\documentclass[twocolumn]{aastex63}

\usepackage{graphicx}	
\usepackage{color}
\usepackage{amsmath}	
\usepackage{amssymb}	

\newcommand{\msun}{M_\odot}
\newcommand{\msunyr}{M_\odot~{\rm yr}^{-1}}
\newcommand{\zsun}{Z_\odot}
\newcommand{\K}{{\rm K}}

\shorttitle{Pulsational instability of very massive stars}
\shortauthors{D. Nakauchi, K. Inayoshi, and K. Omukai}

\begin{document}

\title{Pulsation-driven mass loss from massive stars behind stellar mergers in metal-poor dense clusters}

\author{Daisuke Nakauchi}
\affiliation{Astronomical Institute, Graduate School of Science, Tohoku University, Aoba, Sendai 980-8578, Japan}

\author{Kohei Inayoshi}
\affiliation{Kavli Institute for Astronomy and Astrophysics, Peking University, Beijing 100871, China}

\author{Kazuyuki Omukai}
\affiliation{Astronomical Institute, Graduate School of Science, Tohoku University, Aoba, Sendai 980-8578, Japan}



\begin{abstract}

The recent discovery of high-redshift~($z>6$) supermassive black holes~(SMBH) favors the formation of massive seed BHs in protogalaxies.
One possible scenario is formation of massive stars $\simeq 10^{3}\mbox{-}10^{4}~\msun$ via runaway stellar collisions in a dense cluster, leaving behind massive BHs without significant mass loss.
We study the pulsational instability of massive stars with the zero-age main-sequence~(ZAMS) mass $M_{\rm ZAMS}/\msun = 300\mbox{-}3000$ and metallicity $Z/\zsun = 0\mbox{-}10^{-1}$, and discuss whether or not pulsation-driven mass loss prevents massive BH formation.
In the MS phase, the pulsational instability excited by the $\epsilon$-mechanism grows in $\sim 10^3\ {\rm yrs}$.
As the stellar mass and metallicity increase, the mass-loss rate increases to $\lesssim 10^{-3}\ \msun\ {\rm yr}^{-1}$.
In the red super-giant~(RSG) phase, the instability is excited by the $\kappa$-mechanism operating in the hydrogen ionization zone and grows more rapidly in $\sim 10\ {\rm yrs}$.
The RSG mass-loss rate is almost independent of metallicity and distributes in the range of $\sim 10^{-3}\mbox{-}10^{-2}\ \msun\ {\rm yr}^{-1}$.
Conducting the stellar structure calculations including feedback due to pulsation-driven winds, we find that the stellar models of $M_{\rm ZAMS}/\msun = 300\mbox{-}3000$ can leave behind remnant BHs more massive than $\sim 200\mbox{-}1200\ \msun$.
We conclude that massive merger products can seed monster SMBHs observed at $z>6$.

\end{abstract}

\keywords{stars: evolution, stars: Population III, stars: Population II}



\section{Introduction}\label{sec:intro}

Supermassive black holes~(SMBH) of $10^6\mbox{-}10^{9}\ \msun$ have been discovered in almost all massive galaxies.
The past and ongoing surveys of distant quasars have also revealed more than two hundreds of SMBHs beyond $z>6$, i.e., within 1 Gyr after the Big Bang~\citep[e.g.,][]{Fan2006,Mortlock2011,Venemans2013,Wu2015,Banados2018,Matsuoka2018,Onoue2019}.
The formation of such massive objects within a short timescale is a challenging problem in astrophysics, and requires rapid assembly of massive seed BHs in high-redshift protogalaxies~\citep[e.g.,][]{Volonteri2012, Inayoshi2020}.

A natural formation pathway is to consider the remnant BHs left behind by metal-free Population III~(hereafter Pop III) stars.
Unlike the present-day star-formation where metals and dust grains provide efficient cooling, primordial star-formation proceeds via inefficient H$_2$ cooling, making star-forming clouds significantly warmer than molecular clouds in the present-day universe.
As a result, Pop III stars can be as massive as $10\mbox{-}1000\ \msun$, as recent numerical simulations have found~\citep{Hosokawa2011, Hosokawa2016, Stacy2012, Stacy2016, Hirano2014, Susa2014}.
If Pop III remnant BHs of $10\mbox{-}1000\ \msun$ can grow in mass via the Eddington-limited accretion at 100 per cent duty cycle, then their mass can reach $\sim 10^{9}\ \msun$ within $\sim 1\ {\rm Gyr}$.
In reality, however, radiative feedback prohibits such a high duty-cycle over 6-8 orders of magnitude in mass, so that the growth timescale can be much longer than 1 Gyr~\citep[e.g.,][]{Alvarez2009, Milos2009}.

An alternative scenario is the formation of massive seeds through gravitational collapse of supermassive stars~(SMSs) of $10^5\mbox{-}10^6\ \msun$.
In protogalaxies exposed to strong ultra-violet radiation and experiencing high-density shock-compression, H$_2$ formation is suppressed via photo- and collisional-dissociation, respectively, and star-forming clouds contract almost isothermally with $\sim$ 8000 K via Ly-$\alpha$ emission~\citep[e.g.,][]{Omukai2001, Oh_Haiman2002, Bromm_Loeb2003, Shang2010, Schleicher2010a, Wolcott-Green2011,Inayoshi2012,Agarwal2012,Regan2014,Sugimura2014}.
This H$_2$-free cloud with a mass of $\gtrsim 10^5~\msun$ can collapse monolithically to the central single object without vigorous fragmentation at a high mass-accretion rate of $0.1\mbox{-}1\ M_{\odot}\ {\rm yr}^{-1}$, enabling the embryo protostar to grow to a SMS with $10^5\mbox{-}10^6\ \msun$ within its lifetime of $\sim$ Myr~\citep{Inayoshi_Omukai2014, Becerra2015}.
Even if fragmentation is induced by efficient metal/dust cooling, SMS formation would be assisted by rapid migration of fragments via dynamical friction and disk interaction~\citep{Inayoshi_Haiman2014}.
Recent hydrodynamical simulations by \cite{Chon2020} confirm this in metal-enriched clouds with $Z\lesssim 10^{-3}~\zsun$.
Since an SMS grows at such a high accretion rate as $\gtrsim 0.1~\msunyr$, the stellar envelope is bloated with a surface temperature of $\sim 5000~\K$. 
Therefore, radiative feedback due to stellar ionizing photons does not prevent mass accretion from the collapsing parent cloud~\citep{Hosokawa2013, Schleicher2013, Haemmerle2018}.
Eventually, when the mass reaches $10^5\mbox{-}10^6\ \msun$ \citep[e.g.,][]{Sato1966, Shapiro1983, Umeda2016}, the SMS directly collapses into a remnant BH with a similar mass by the general-relativistic instability, nearly regardless of stellar rotation and nuclear fusion activated during the collapse phase~\citep{Shibata2002, Uchida2017}\footnote{
For an SMS with high metallicities, runaway nuclear fusion might cause a very energetic explosion~\citep{Montero2012}.}.
The formation of massive seeds gives them a head start to be $\sim 10^{9}\ \msun$, shortening the required growth timescale.

The third possible channel is through very massive stars with $\sim 10^{3}\mbox{-}10^4~\msun$ via runaway stellar collisions in dense clusters.
When the stellar density is sufficiently high ($\gtrsim 10^5~\msun~{\rm pc}^{-3}$), direct collisions between stars can take place quickly within a timescale shorter than their lifetime~\citep{Katz2015, Yajima2016, Sakurai2017}, leaving a single massive star.
In slightly metal-enriched clouds~($\sim 10^{-5}\mbox{-}10^{-3}~\zsun$) of protogalaxies, dense clusters with stellar masses of $\sim 10^5~\msun$ and half-mass radii of $\sim$ 1 pc would be formed by fragmentation via metal and dust cooling~\citep[e.g.,][]{Omukai2008,Devecchi2009}.
Using $N$-body simulations, \cite{Katz2015}, \cite{Sakurai2017}, and \cite{Reinoso2018} show that more massive stars segregate to the center within their lifetime~(a few Myr) and start collisions with ambient lower-mass stars in a runaway fashion, forming a very massive star of $\gtrsim 1000\ \msun$.
We note that in the presence of mass accretion onto the central stars, the bloated stellar radii~(i.e., larger cross-sections) increase the stellar-collision rate by up to an order of magnitude, enabling the formation of more massive stars of $\gtrsim 10^4\ \msun$~\citep{Boekholt2018,Seguel2020,Tagawa2020}.

Even if very massive stars successfully form, it is not ensured that they can form massive remnant BHs of the same masses before losing significant mass by pulsation- and radiation-driven winds.
In the case of SMS formation with rapid mass accretion~($\gtrsim 0.1\ M_{\odot}\ {\rm yr}^{-1}$), the mass loss associated with both pulsation- and radiation-driven winds is too weak to prevent the stellar growth~\citep{Inayoshi2013,Nakauchi2017}.
On the other hand, in the runaway collision scenario without gas accretion, stars gain mass more episodically via stellar collisions alone.
The post-merger product contracts its radius in the Kelvin-Helmholtz timescale and evolves into the main-sequence (MS) structure during two successive mergers.
If the mass-loss rate exceeds the mass-supplying rate by stellar collisions, the stellar growth is prohibited and the remnant BH mass could be much lower than originally thought.
Therefore, it is crucial in the runaway collision scenario to investigate the stability of very massive stars and estimate the mass-loss rate for various situations.

Previous authors \citep{Baraffe2001, Sonoi2012, Shiode2012,Inayoshi2013} conducted the linear stability analyses for zero-metallicity massive stars with $100\mbox{-}3000\ \msun$ and found that while these stars are pulsationally unstable in the MS phase, the pulsation-driven mass-loss rate~($\simeq 10^{-6}\mbox{-}10^{-4}~\msunyr$) is too small to affect the stellar evolution.
\cite{Baraffe2001} and \cite{Shiode2012} also studied the instability of $\sim 100\ \msun$ stars for various metallicities of $Z/\zsun = 10^{-4}, 10^{-3}, 10^{-2}$, and $10^{-1}$, and found the mass-loss rate to be higher with metallicity.
However, all the previous works have not investigated the stability in the post-MS stages extensively~(note that \cite{Heger1997} and \cite{Moriya2015} studied the stability of red-supergiants~(RSG) both at solar and zero metallicities and found they tend to be more unstable with increasing mass).
Extending the parameter space for the stellar mass and metallicity, in this paper, we study the pulsational instability of massive stars in both the MS and post-MS stages.
We also calculate the stellar evolution by accounting for feedback due to pulsation-driven mass loss, and discuss the validity of the runaway-stellar-merger scenario as a massive-BH-seeding mechanism.

The rest of the paper is organized as follows.
In Section \ref{sec:method}, we describe the method for the stellar evolution calculation and the linear stability analysis.
Pulsationally unstable models are shown in Section \ref{subsec:star_model}, with the discussion of the growth rate, excitation mechanism, and mass loss due to instability.
In Section \ref{subsec:mass_loss}, to estimate the final mass of stars, the stellar structure evolution is calculated by including the backreaction of pulsation-driven mass loss.
In Section \ref{sec:summary}, after the brief summary, we discuss the implications and uncertainties of our results.

\vspace{10mm}
\section{Method}\label{sec:method}

\subsection{Mechanical and thermal equilibrium stellar model}\label{subsec:method_star}

The stellar models at various evolutionary stages are computed by the Modules for Experiments in Stellar Astrophysics~\citep[MESA release 12115;][]{Paxton2011,Paxton2013,Paxton2015,Paxton2018,Paxton2019}.
We neglect rotation and mass loss by radiation-driven winds.
Note that pulsation-driven mass loss is taken into account later~(see Section \ref{subsec:method_mass_loss}).
The onset of convection is determined by the Ledoux criterion, and the convective energy flux is calculated by the mixing length theory with a mixing length of $\alpha_{\rm MLT} = 1.8$ local pressure scale-height.
At the boundaries of convective regions, the convective overshoot is assumed to occur in the extent of $f_{\rm ov} = 0.015$ pressure scale-height.
In addition, the semi-convective mixing is considered with the formalism of \cite{Langer1983}, where the dimensionless efficiency parameter is set to $\alpha_{\rm sv} = 10$.

We consider 24 stellar models with different masses in the zero-age main-sequence~(ZAMS) stage of $M_{\rm ZAMS}/\msun = 300, 500, 750, 1000$, and $3000$, and metallicities of $Z/\zsun = 0, 10^{-4}, 10^{-3}, 10^{-2}$, and $10^{-1}$, except the case with $M_{\rm ZAMS} = 3000\ \msun$ and $Z = 10^{-1}\ \zsun$.
The evolutionary track in each model is calculated from the ZAMS stage to a post-MS stage between the hydrogen depleting and helium depleted times~(the corresponding stellar age is $\sim 2.0\mbox{-}2.5$ Myr).
After helium depletion, the residual lifetime is so short~\citep[$<$ 100 yrs;][]{Woosley2002} that the subsequent core evolution does not affect the total mass loaded into pulsation-driven winds.
Among the models, the evolution until helium depletion is successfully followed for all the four models with $Z/\zsun = 10^{-1}$.
Due to difficulties in numerical convergence, for the four models of $(M_{\rm ZAMS}/\msun, Z/\zsun) = (1000, 10^{-4}), (750, 10^{-3}), (1000, 10^{-3})$, and ($3000, 10^{-2})$, the simulations are terminated before hydrogen depletion. 
In the remaining 16 models, the computations are stopped before helium depletion.

\vspace{5mm}

\subsection{Stability analysis}\label{subsec:method_stability}

The instability of radial pulsation modes is considered for the stellar models computed in the previous subsection.
Each physical quantity $Q$ is perturbed in the Lagrangian form as $\delta Q(r,t) \equiv Q(r+\xi_r, t) - Q_0(r)$, where $Q_0(r)$ is the equilibrium value and $\xi_r$ the radial displacement from the equilibrium position.
Hereafter, the subscript `0' is omitted.
By separating the temporal dependence from the spatial one as $\delta Q(r,t) = \delta Q(r) \exp(i \sigma t)$, where $\sigma~(= \sigma_{\rm R}+i\sigma_{\rm I})$ is the eigen-frequency ($\sigma_{\rm R}$ is the frequency of the pulsation and $\sigma_{\rm I}$ is the growing or damping rate of the pulsation depending on its sign), the linearized perturbation equations for continuity, motion, energy and radiative energy transport are written, respectively, as follows:
\begin{equation}
\frac{{\rm d}}{{\rm d}M_r} \xi_r = - \frac{1}{4 \pi r^2 \rho} \left(2 \frac{\xi_r}{r} + \frac{\delta \rho}{\rho}\right),
\label{eq:perturb_xi}
\end{equation}
\begin{equation}
\frac{{\rm d}}{{\rm d}M_r} \delta P = \left(\frac{\sigma^2}{4 \pi r} + \frac{G M_r}{\pi r^4} \right)\frac{\xi_r}{r},
\label{eq:perturb_P}
\end{equation}
\begin{equation}
\frac{{\rm d}}{{\rm d} M_r} (\delta L_{\rm rad} + \delta L_{\rm conv})= \delta \epsilon_{\rm nuc} - i \sigma T \delta S,
\label{eq:perturb_S}
\end{equation}
\begin{equation}
\frac{{\rm d}}{{\rm d} M_r}  \delta T = \frac{{\rm d} T}{{\rm d} M_r}\left(\frac{\delta L_{\rm rad}}{L_{\rm rad}} - 3 \frac{\delta T}{T} - 4 \frac{\xi_r}{r} + \frac{\delta \kappa}{\kappa} \right),
\label{eq:perturb_Lrad}
\end{equation}
where $\delta \rho, \delta P, \delta L_{\rm rad}, \delta L_{\rm conv}, \delta \epsilon_{\rm nuc}, \delta S, \delta T$, and $\delta \kappa$ indicate the Lagrangian perturbation of density, pressure, radiative and convective luminosity, nuclear energy generation rate, entropy, temperature, and opacity, respectively, and $M_r$ the enclosed mass.
Owing to the lack of understanding of the interaction between convective motion and pulsation, the perturbation of convective luminosity is often neglected \citep[the so-called frozen-in approximation, e.g.,][]{Unno1989}.
Hereafter, we adopt this prescription for simplicity and set $\delta L_{\rm conv} = 0$ in Eq. \eqref{eq:perturb_S}.

We impose four boundary conditions, i.e., two at the center and the other two at the surface, to solve Eqs. \eqref{eq:perturb_xi}-\eqref{eq:perturb_Lrad}.
At the center~($M_r = 0$), since the radial displacement is zero, and the central region is sufficiently adiabatic,
\begin{equation}
\xi_r = 0,\ \text{and}\ \delta S = 0.
\end{equation}
At the surface~($M_r = M_\ast$), we impose the regularity condition of Eq. \eqref{eq:perturb_P},
\begin{equation}
\frac{\delta P}{P} = - \left(\sigma^2 \frac{R_\ast^3}{GM_\ast} +4 \right) \frac{\xi_r}{R_\ast},
\label{eq:bc_surf_1}
\end{equation}
and the photospheric condition
\begin{equation}
\frac{\delta L_{\rm rad}}{L_{\rm rad}} - 2 \frac{\xi_r}{R_\ast} - 4 \frac{\delta T}{T} = 0.
\label{eq:bc_surf_2}
\end{equation}

Eqs. \eqref{eq:perturb_xi}-\eqref{eq:perturb_Lrad} with the above four boundary conditions become a two-point boundary value problem with an eigenvalue $\sigma~(= \sigma_{\rm R} + i \sigma_{\rm I})$.
We use the numerical code developed by \cite{Inayoshi2013}, where the relaxation method described in Section 18.2 of \cite{Unno1989} is adopted.
Following \cite{Inayoshi2013}, we focus on the fundamental mode (i.e., no nodes in the eigenfunction).
The pulsation period and growth rate are calculated by $\Pi = 2 \pi/\sigma_{\rm R}$ and $t_{\rm grow} = -\sigma_{\rm I}^{-1}$, respectively.
The radial pulsation mode is unstable~(or stable) when $\sigma_{\rm I}^{-1} < 0$~($ >0$, respectively).

The excitation and damping of a pulsation mode can be understood by calculating the work integral $W(M_r)$, defined as the change of the pulsation energy within the enclosed mass $M_r$ over a pulsation cycle~\citep{Unno1989}:
\begin{equation}
W(M_r) = \frac{\pi}{\sigma_{\rm R}} \int_0^{M_r} \frac{\delta T^\ast}{T} \left(\delta \epsilon_{\rm nuc} - \frac{{\rm d} \delta L_{\rm rad}}{{\rm d} M_r} \right)\,dM_r. 
\label{eq:work_integral}
\end{equation}
The first term in the integration presents the heat obtained from the nuclear energy generation in one cycle and is rewritten as
\begin{equation}
\frac{\delta T^\ast}{T} \delta \epsilon_{\rm nuc} = \left|\frac{\delta T}{T}\right|^2 \left[\epsilon_T + \frac{\epsilon_\rho}{\Gamma_3 -1}\right] \epsilon_{\rm nuc},
\label{eq:epsilon}
\end{equation}
where $\epsilon_T=\left(\partial \ln \epsilon/\partial \ln T\right)_\rho$ and $\epsilon_\rho=\left(\partial \ln \epsilon/\partial \ln \rho\right)_T$.
This term is always positive and shows the destabilization by the so-called $\epsilon$-mechanism.
On the other hand, the second term characterizes the pulsation dumping due to radiative diffusion and excitation due to absorption of the radiative flux in the surface layer where the opacity changes remarkably.
By maintaining only the dominant terms, the second term is written as
\begin{equation}
\frac{\delta T^\ast}{T} \left(- \frac{{\rm d} \delta L_{\rm rad}}{{\rm d} M_r} \right) \approx  L_{\rm rad} \left|\frac{\delta T}{T}\right|^2 \frac{{\rm d}}{{\rm d} M_r} \left(\kappa_T + \frac{\kappa_\rho}{\Gamma_3 -1} \right),   
\label{eq:2ndterm}
\end{equation}
where $\kappa_T =\left(\partial \ln \kappa/\partial \ln T\right)_\rho$ and $\kappa_\rho=\left(\partial \ln \kappa/\partial \ln \rho\right)_T$.
Therefore, only when the opacity satisfies
\begin{equation}
\frac{{\rm d}}{{\rm d} M_r} \left(\kappa_T + \frac{\kappa_\rho}{\Gamma_3 -1} \right) > 0,
\label{eq:kappa}
\end{equation}
stars are destabilized~(the so-called $\kappa$-mechanism).

When the perturbation grows slowly in an oscillation period (i.e., $|\sigma_{\rm I}| \ll \sigma_{\rm R}$, $\sigma_{\rm I}$), the work integral is related to the growth (or damping) rate of the pulsation per single period as
\begin{equation}
\eta \equiv - \frac{\sigma_{\rm I}}{\sigma_{\rm R}} = \frac{W(M_\ast)}{4 \pi E_{\rm puls}},
\label{eq:eta}
\end{equation}
where $E_{\rm puls}$ is the pulsation energy defined by
\begin{equation}
E_{\rm puls} = \frac{1}{2} \sigma_{\rm R}^2 \int_0^{M_\ast} |\xi_r|^2\ dM_r.
\label{eq:E_puls}
\end{equation}
Therefore, when $W(M_\ast) > 0$~(or $<0$), $\sigma_{\rm I} < 0$~($>0$) and stars are pulsationally unstable~(stable, respectively).

In unstable models, the pulsation amplitude grows up to the non-linear regime in $t_{\rm grow}$.
Then, the pulsation energy becomes large enough for the surface materials to escape from the star~\citep{Appenzeller1970a,Appenzeller1970b,Yadav2018}.
Following \cite{Baraffe2001}, the mass-loss rate is estimated by assuming that all the pulsation energy is used to lift the surface materials against gravity as:
\begin{equation}
\frac{1}{2} \dot{M}_{\rm puls} v_{\rm esc}^2 \sim L_{\rm puls} = \frac{{\rm d}E_{\rm puls}}{{\rm d} t} = 2 |\sigma_{\rm I}| E_{\rm puls}, 
\label{eq:mdot_puls}
\end{equation}
where $v_{\rm esc} = (G M_\ast/R_\ast)^{1/2}$ is the escape velocity.
When we calculate the pulsation energy $E_{\rm puls}$ in Eq. \eqref{eq:E_puls}, the radial displacement $\xi_r$ is obtained by extrapolating the solution of the perturbation equations into the non-linear regime.
When the mass loss sets in, the oscillation speed at the surface is equal to the sound velocity there~($c_{\rm s, \ast}$), so that the displacement at the surface is determined from $\xi_r(M_\ast)=c_{\rm s, \ast}/\sigma_{\rm R}$.

From Eqs. \eqref{eq:E_puls} and \eqref{eq:mdot_puls}, we can discuss how the pulsation energy and mass-loss rate depend on the stellar parameters.
First, from Eq. \eqref{eq:E_puls}, the pulsation energy can be represented as
\begin{equation}
E_{\rm puls} \sim \sigma_{\rm R}^2 \int_0^{M_\ast} |\xi_r|^2 dM_r \propto M_{\ast} T_{\rm eff} \int_0^{1} |\tilde{\xi}_r|^2 dq_r,
\label{eq:E_puls_para}
\end{equation}
where we used $q_r = M_r/ M_\ast$, $c_{\rm s, \ast} \propto T_{\rm eff}^{1/2}$, and the non-dimensional form of the radial displacement $\tilde{\xi}_r$ defined by $\xi_r  = \tilde{\xi}_r c_{\rm s, \ast}/\sigma_{\rm R}$.
While the integral part changes with the stellar structure~(e.g., if a star has an extended envelope or not), the integral part hardly depends on the stellar mass.
Therefore, the pulsation energy is found to be proportional to the stellar mass.
Next, from Eqs. \eqref{eq:mdot_puls} and \eqref{eq:E_puls_para}, the mass-loss rate can be represented as
\begin{align}
\dot{M}_{\rm puls} \sim& \frac{|\sigma_{\rm I}| E_{\rm puls}}{v_{\rm esc}^2} \propto |\sigma_{\rm I}| R_{\ast}c_{\rm s, \ast}^2 \int_0^{1} |\tilde{\xi}_r|^2 dq_r \nonumber \\
&\propto |\sigma_{\rm I}| M_{\ast}^{1/2} T_{\rm eff}^{-1} \int_0^{1} |\tilde{\xi}_r|^2 dq_r,
\label{eq:mdot_para}
\end{align}
where we used $R_{\ast} \propto L_{\ast}^{1/2} T_{\rm eff}^{-2} \propto M_{\ast}^{1/2} T_{\rm eff}^{-2}$, as the luminosity is close to the Eddington limit.
Since $|\sigma_{\rm I}|$ is found to vary with the stellar mass, effective temperature, and metallicity~(see Figures \ref{fig:growth_rate_z2} and \ref{fig:growth_rate_m1e3} below), the mass-loss rate can be described as a function of $M_{\ast}$, $T_{\rm eff}$, and $Z$.

\vspace{5mm}

\subsection{Evolution calculation with pulsation-driven mass loss}\label{subsec:method_mass_loss}

To examine how the pulsational instability affects the evolutionary tracks as well as the final stellar mass, stellar evolution is calculated by considering feedback due to pulsation-driven mass loss.
For each metallicity, we adopt the mass-loss rate obtained from the stability analysis~(see more details in Section \ref{subsec:star_model}).
Pulsation-driven mass loss is implemented by turning on the {\tt use\_other\_wind} control option and by adding customized subroutines to the MESA code.
Other than the mass loss prescription, we set the same model parameters as described in Section  \ref{subsec:method_star}.
The input files~(inlists) and source files to reproduce our results are provided at \url{http://cococubed.asu.edu/mesa_market/inlists.html}.

With mass loss, the stellar evolution calculations are conducted until helium depletion for 11 models; namely, all the five models with $Z/\zsun = 10^{-2}$ and the models of $M_{\rm ZAMS}/\msun = 750$ and $1000$ with $Z/\zsun = 10^{-4}, 10^{-3}$, and $10^{-1}$.
Due to the issues of numerical convergence, for the four models of $(M_{\rm ZAMS}/\msun, Z/\zsun) = (750, 0), (1000, 0), (500, 10^{-4})$, and $(500, 10^{-3})$, the simulations are terminated just before hydrogen depletion.
For the remaining 8 cases, the evolutionary tracks are calculated until the beginning of helium depletion.
In summary, the simulation results are shown for the above 23 stellar models in Section \ref{subsec:mass_loss}~(note that the $(M_{\rm ZAMS}/\msun, Z/\zsun)=(300, 10^{-1})$ model does not reach the hydrogen exhaustion stage due to numerical issues).

\vspace{10mm}

\section{Results}\label{sec:result}

We here show the results of the stellar evolution calculations and linear stability analysis for massive star models.
In Section \ref{subsec:star_model}, we first construct stellar models without taking into account mass loss and estimate the mass-loss rate driven by stellar pulsation.
As shown below, the mass-loss rate depends on the stellar mass and surface temperature.
In Section \ref{subsec:mass_loss}, we show the results of stellar evolution models taking into account pulsation-driven mass loss.

\vspace{5mm}

\subsection{Stellar models and instability}\label{subsec:star_model}

\begin{figure*}
\begin{center}
\begin{tabular}{cc}
{\includegraphics[scale=1.1]{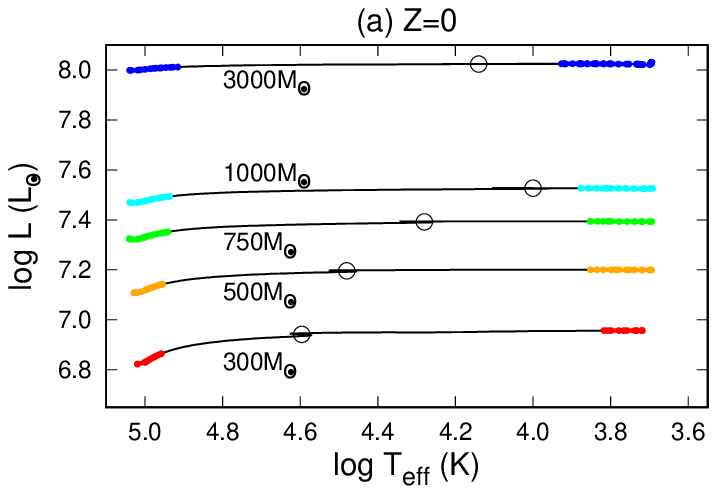}}
{\includegraphics[scale=1.1]{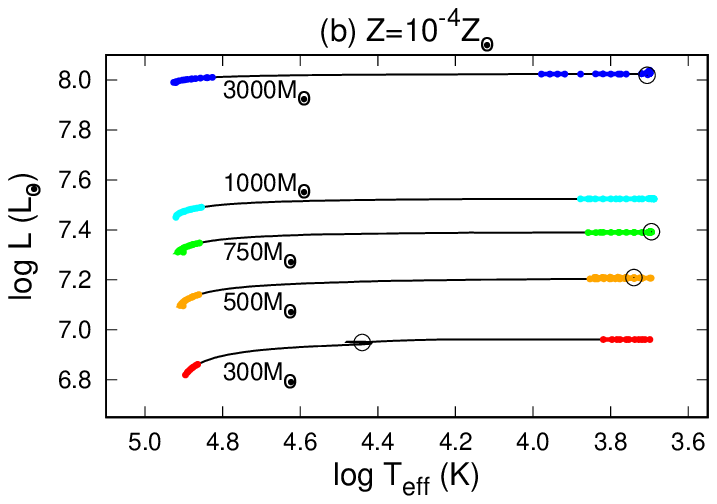}}\\
{\includegraphics[scale=1.1]{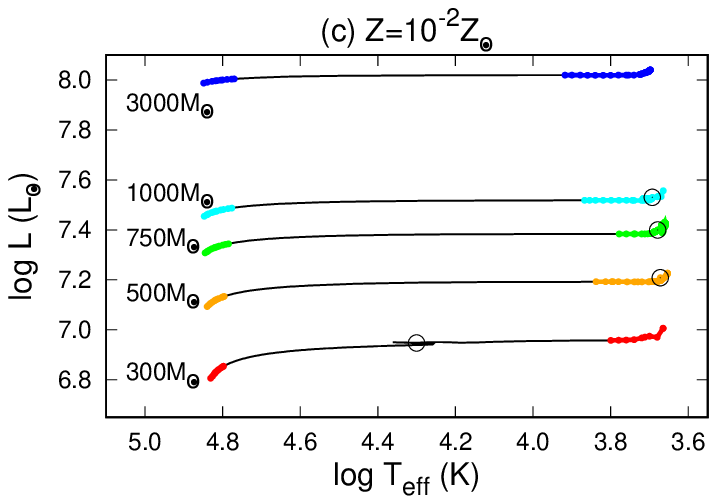}}
{\includegraphics[scale=1.1]{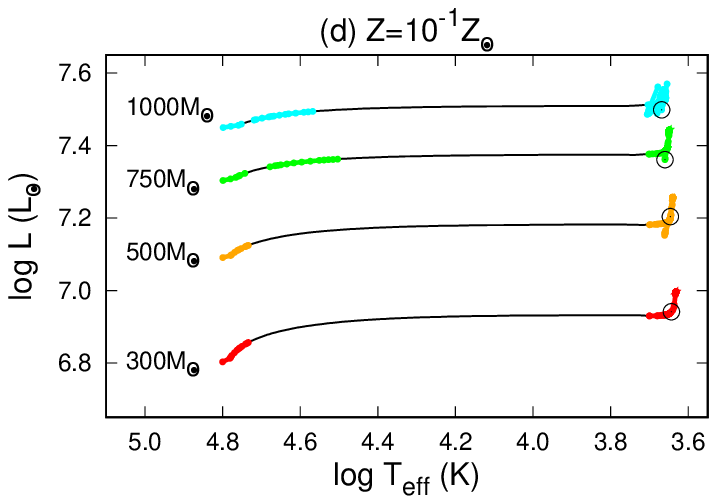}}\\
\end{tabular}
\caption{Stellar evolutionary tracks in the HR diagram. Pulsationally unstable models are denoted by colored filled circles.
Panels (a)-(d) show the results for $Z/\zsun = 0, 10^{-4}, 10^{-2}$, and $10^{-1}$, respectively.
In each panel, evolutionary tracks are shown for the models with $M_{\rm ZAMS}/\msun= 300$, $500$, $750$, $1000$, and $3000$ from the bottom to the top.
Open circles represent the epoch of hydrogen depletion in the core.
All the stellar models are pulsationally unstable in the early MS stage and in the RSG stage.}
\label{fig:hrd_unstable}
\end{center}
\end{figure*}

In Figure \ref{fig:hrd_unstable}, we show the evolutionary tracks of stellar models with different metallicities, $Z/\zsun =$ (a) $0$, (b) $10^{-4}$, (c) $10^{-2}$, and (d) $10^{-1}$, in the Herzsprung-Russell (HR) diagram.
Each panel presents the cases with different stellar masses at $300 \leq M_{\rm ZAMS}/\msun \leq 3000$, the values of which are denoted by the numbers in the figure.
Open circles represent the epochs of hydrogen depletion in the stellar core.
For each case, colored points show the stellar models that are unstable against the radial linear perturbation and the consecutive unstable models are connected with the bold lines.

All the stellar models begin to evolve from the leftmost side of the HR diagram in their ZAMS stages.
For each metallicity case, we find the mass-luminosity and mass-radius relations to be $L_{\rm ZAMS} \propto M_{\rm ZAMS}$ and $R_{\rm ZAMS} \propto M_{\rm ZAMS}^{0.5}$, respectively.
Therefore, the surface temperature is almost independent of the stellar mass, $T_{\rm eff} \propto L_{\rm ZAMS} R_{\rm ZAMS}^{-2} \propto M_{\rm ZAMS}^0$, and distributes in the range of $\simeq 10^{4.8}\mbox{-}10^5~\K$~(see Figure \ref{fig:hrd_unstable}).
On the other hand, for a fixed stellar mass, lower-metallicity models are found to show higher surface temperatures and are more compact.
This is because, with smaller amount of carbon, metal-poor stars should keep the core temperature higher to gain sufficient nuclear energy via CN cycle and support the entire stellar structure.
In the ZAMS phase, all the models are pulsationally unstable owing to the nuclear energy generation in the cores~(the so-called $\epsilon$ mechanism).

As hydrogen is consumed in the core, the stellar envelope expands with the surface temperature decreasing monotonically~(see Figure \ref{fig:hrd_unstable}).
All the models are stabilized before hydrogen core depletion.
However, when those stars evolve into RSG~($T_{\rm eff} < 10^4~\K$), they become unstable again.
The instability in the RSG phase is due to the blocking of radiative flux at the ionization layers of atoms~(the so-called $\kappa$ mechanism; see more details below).

\begin{figure*}
\begin{center}
\begin{tabular}{cc}
{\includegraphics[scale=1.1]{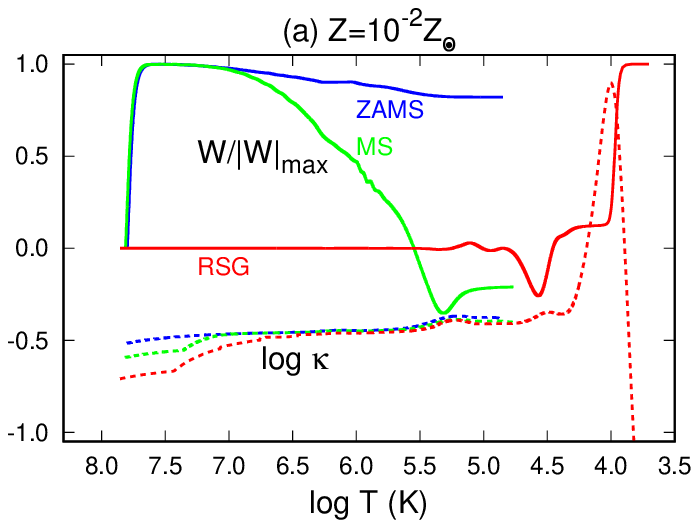}}
{\includegraphics[scale=1.1]{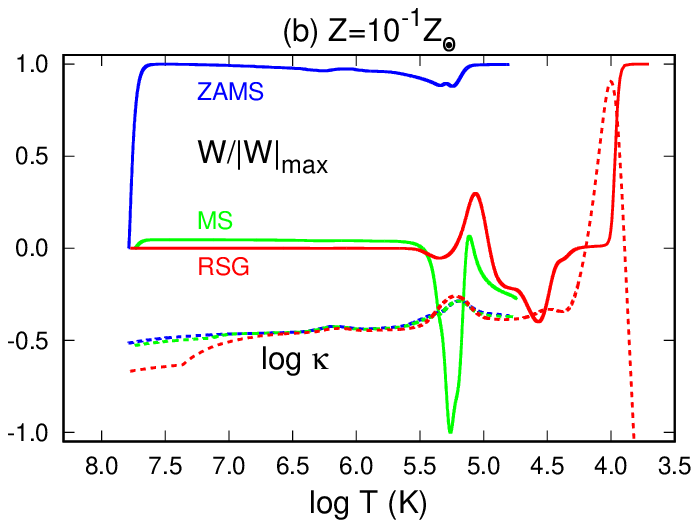}}\\
\end{tabular}
\caption{Radial distribution of the work integral $W$ normalized by the maximum value $|W|_{\rm max}$~(solid lines), and opacity~(dashed lines) as a function of the stellar interior temperature in the case of $M_{\rm ZAMS} = 1000\ \msun$.
In panels (a) and (b), the cases with $Z/\zsun = 10^{-2}$ and $10^{-1}$ are shown.
In each panel, the colored curves correspond to the models in the ZAMS~(blue), MS~(green), and RSG~(red), respectively.
In the ZAMS model, the pulsational instability is excited by nuclear burning in the core~($W>0$).
As the star evolves, the star is stabilized via radiative diffusion in the outer-most layer~($W<0$).
In the RSG model, the star becomes destabilized in the hydrogen ionization layer at $\log T \sim 4.0$~($W>0$), almost independently of metallicity.}
\label{fig:work_integral}
\end{center}
\end{figure*}

In Figure \ref{fig:work_integral}, we present the radial distribution of the work integral $W$ normalized by the maximum value $|W|_{\rm max}$~(solid), and opacity~(dashed) as a function of temperature for the $M_{\rm ZAMS} = 1000\ \msun$ models with $Z/\zsun = 10^{-2}$~(panel a) and $10^{-1}$~(panel b).
Different curves correspond to the cases of the ZAMS~(blue), MS~(green), and RSG~(red) phase, respectively.
Note that the results with $Z/\zsun \leq 10^{-3}$ are quantitatively similar to those with $Z/\zsun = 10^{-2}$, and thus are not shown.
A positive (negative) value of $W$ at the surface indicates that the stellar model is unstable (stable) against linear perturbations.

In both the ZAMS and MS models (blue and green curves), the work integral increases in the core region where the instability is excited by nuclear burning.
In the MS models, the work integral drops to a negative value at the outermost layers of $\log T \leq 5.5$, because the pulsation energy excited in the core damps via radiative diffusion in the envelope where the opacity varies smoothly ~\citep[e.g.,][]{Schwarzschild1959,Baraffe2001}.
In the higher-metallicity case ($Z/\zsun = 10^{-1}$;  panel b), the $\kappa$-mechanism caused by the opacity bump at $\log T \sim 5.2$ due to the bound-bound transitions of iron elements contributes to the excitation of stellar pulsation.
However, it turns out that the two MS models become stable because of pulsation damping by radiative diffusion at the outer-most layers.

The RSG models~(red curves) with $T_{\rm eff} \sim 10^{3.7}~\K$ have bloated envelopes, where the opacity changes remarkably in the ionization layers of hydrogen at $\log T \sim 4.0$, and thus pulsation is excited owing to the absorption of radiative flux by hydrogen bound-free transition.
We note that in the pulsation-driving zone, the radial profiles of opacity and work integral are almost identical among all the cases with different metallicities.
This implies that the strength of pulsations in the RSG hardly depends on the stellar metallicity~(see the discussion below).

\begin{figure}
\begin{center}
\begin{tabular}{c}
{\includegraphics[scale=1.1]{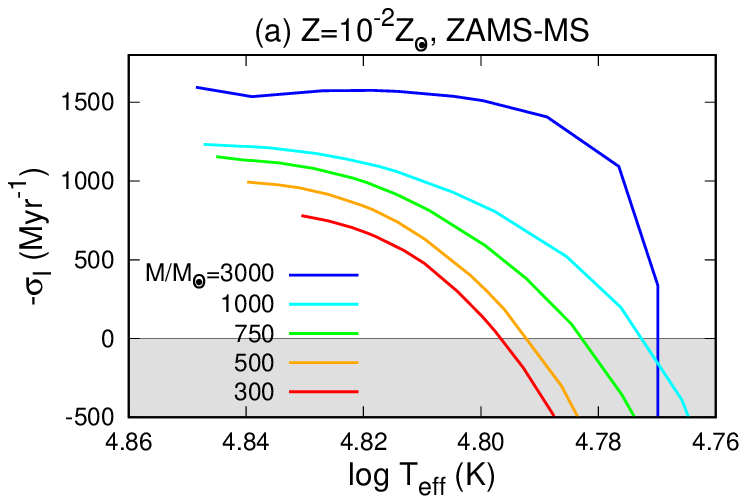}}\\
{\includegraphics[scale=1.1]{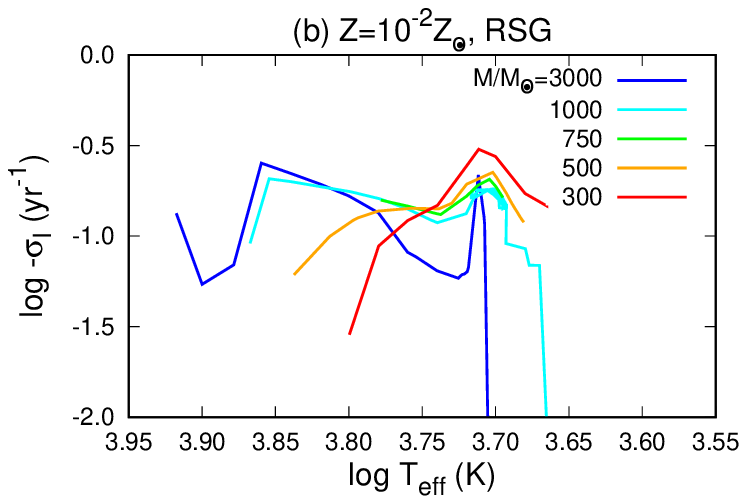}}
\end{tabular}
\caption{The growth rate of pulsational instability as a function of the surface temperature in the ZAMS-MS phase~(panel a) and in the RSG phase~(panel b) for the models with $Z =10^{-2}\ \zsun$ and various stellar masses indicated in the legend. In the grey-shaded regions, $\sigma_{\rm I} > 0$ and stars are stable.
Figure \ref{fig:growth_rate_z2} shows the growth rate of instability~($-\sigma_{\rm I}^{-1}$) as a function of the surface temperature, in the MS (panel a) and RSG~(panel b) phase for different stellar masses of $300 \leq M_{\rm ZAMS}/\msun \leq 3000$ with $Z=10^{-2}~\zsun$.
Note that stars are pulsationally stable in the grey-shaded regions~($\sigma_{\rm I} > 0$).}
\label{fig:growth_rate_z2}
\end{center}
\end{figure}

\begin{figure}
\begin{center}
\begin{tabular}{c}
{\includegraphics[scale=1.1]{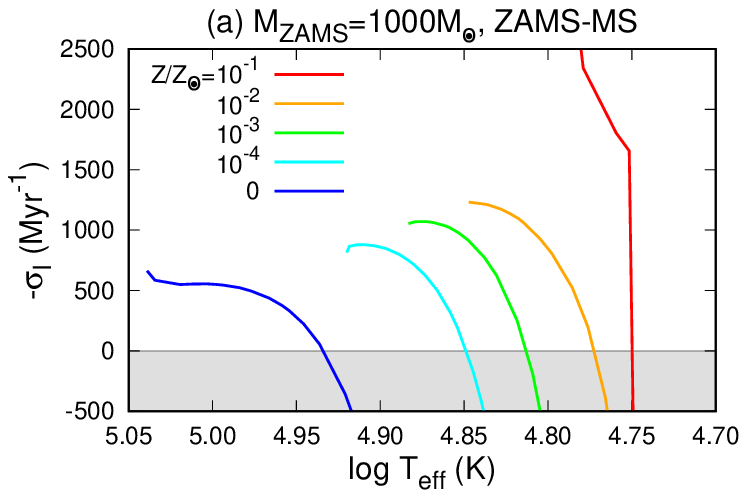}}\\
{\includegraphics[scale=1.1]{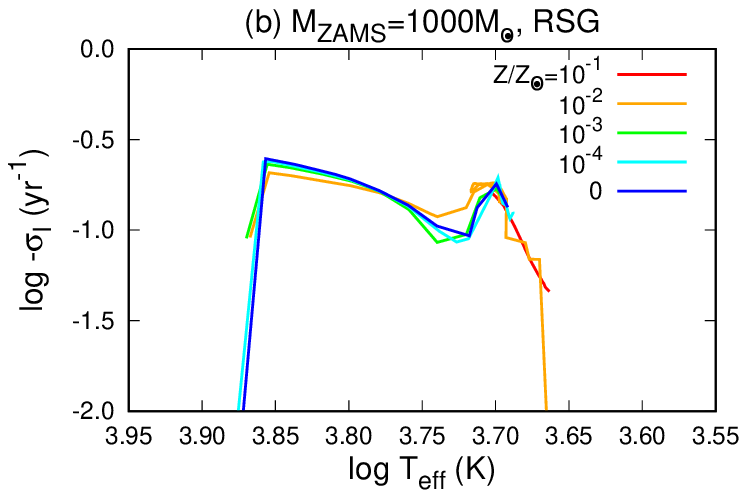}}
\end{tabular}
\caption{The same as Figure \ref{fig:growth_rate_z2} but showing the models with $M_{\rm ZAMS} =1000\ \msun$ and various metallicities indicated in the legend.
Figure \ref{fig:growth_rate_m1e3} presents the metallicity dependence of the growth rate for the $M_{\rm ZAMS} = 1000~\msun$ models.}
\label{fig:growth_rate_m1e3}
\end{center}
\end{figure}

Figure \ref{fig:growth_rate_z2} shows the growth rate of instability~($-\sigma_{\rm I}^{-1}$) as a function of the surface temperature, in the MS (panel a) and RSG~(panel b) phase for different stellar masses of $300 \leq M_{\rm ZAMS}/\msun \leq 3000$ with $Z=10^{-2}~\zsun$.
Note that stars are pulsationally stable in the grey-shaded regions ($\sigma_{\rm I} > 0$).
All the models are unstable in the early MS phase, and are finally stabilized as they evolve off the ZAMS phase.
This is because stellar pulsation excited by the $\epsilon$-mechanism in the core is damped by radiative diffusion in the envelope, as discussed in Figure \ref{fig:work_integral}.
In the early MS phase~(panel a), the growth time of instability is as short as $t_{\rm grow} \sim 10^{3}$ yr, so that stellar pulsation grows into a non-linear regime within the stellar lifetime of $\sim$ Myr.
The growth rate increases for more massive stars because $\sigma_{\rm I}~(\propto \sigma_{\rm R}^{-2} \propto R_{\rm ZAMS}^3/M_{\rm ZAMS})\propto M_{\rm ZAMS}^{1/2}$ from Eq. \eqref{eq:eta}.
In the RSG phase~(panel b), stars are destabilized by the $\kappa$-mechanism operating in the ionization zones of hydrogen, as seen in Figure \ref{fig:work_integral}.
The characteristic growth rate in the RSG phase is as high as $|\sigma_{\rm I}| \sim 0.1\ {\rm yr}^{-1}$, which is two orders of magnitude higher than those in the MS phase.
This fact indicates that the instability grows rapidly in the RSG phase and leads to vigorous mass ejection even in such a short stage.

Figure \ref{fig:growth_rate_m1e3} presents the metallicity dependence of the growth rate for the $M_{\rm ZAMS} = 1000~\msun$ models.
In all the metallicity cases, the overall evolutionary behavior of the growth rate is qualitatively similar to those seen in Figure \ref{fig:growth_rate_z2}.
In the MS phase~(panel a), the growth rate decreases with lower metallicities.
The primary reason is that metal-poor stars have smaller radii~(for a fixed mass) and the growth rate follows $\sigma_{\rm I} \propto R_{\rm ZAMS}^3$.
In addition, the growth rate in the MS stage is proportional to $\epsilon_T$, which is smaller at higher temperature cores of metal-poor stars~\citep[see Fig. 18.8 in][]{Kippenhahn2012}.
In contrast, the growth rate in the RSG phase~(panel b) is almost independent of metallicity, reflecting the fact that the radial structures of work integral and opacity are quite similar among stellar models with different metallicities.

\begin{figure*}
\begin{center}
\begin{tabular}{cc}
{\includegraphics[scale=1.2]{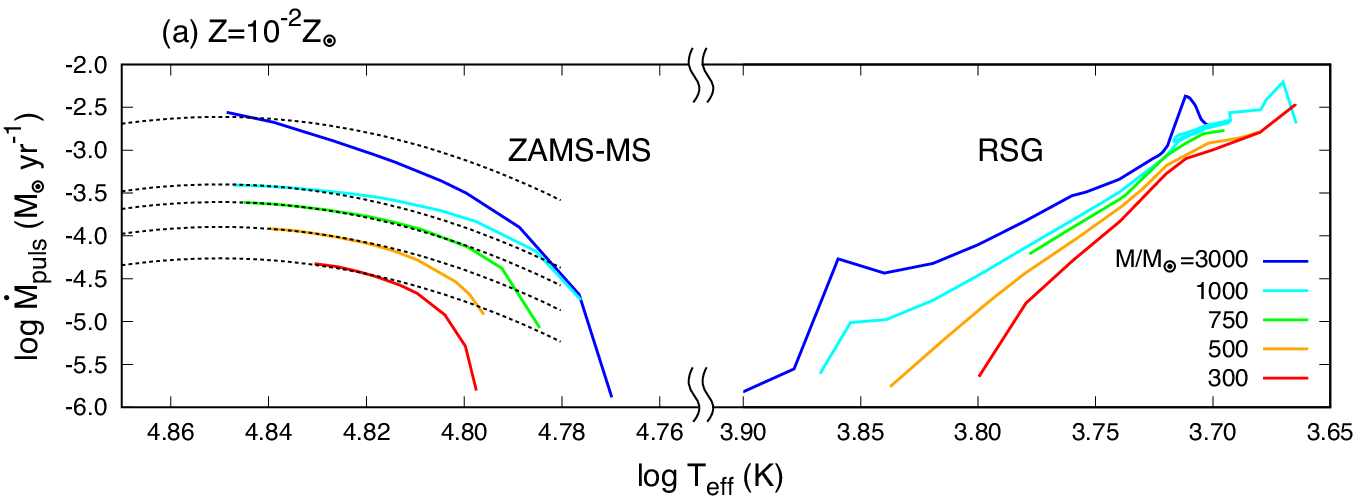}}\\
{\includegraphics[scale=1.2]{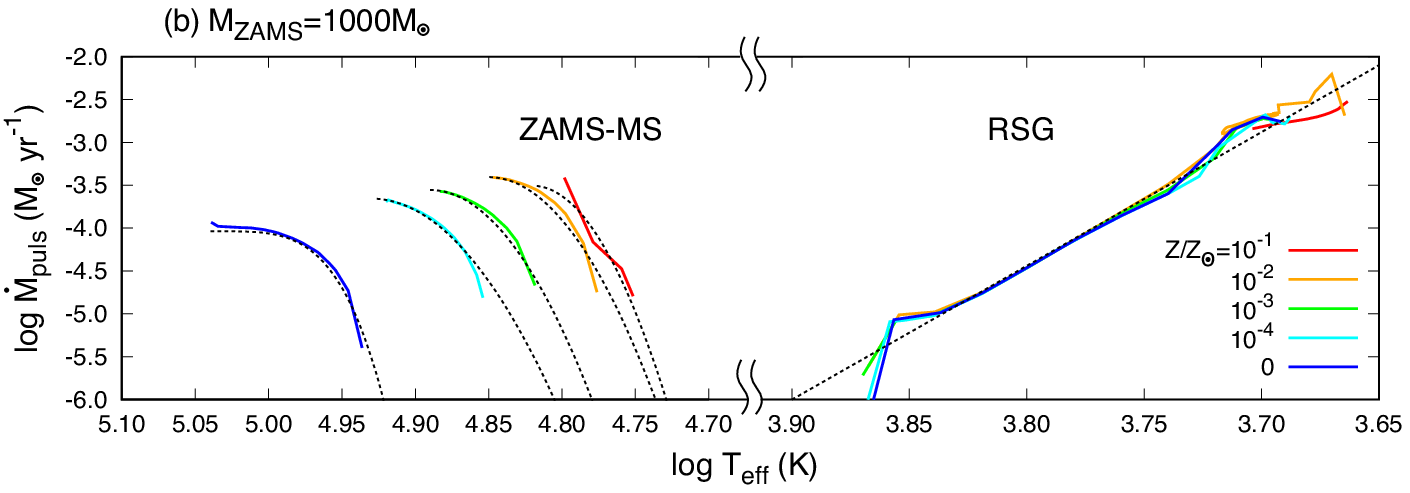}}\\
\end{tabular}
\caption{The mass-loss rate of pulsational instability as a function of the surface temperature for the same models with Figures \ref{fig:growth_rate_z2}~(panel a) and \ref{fig:growth_rate_m1e3}~(panel b).
While the mass-loss rate in the MS stage increases with the stellar mass and metallicity, it converges in the RSG stage to $\gtrsim 10^{-3}~\msunyr$ at $\log T < 3.7$, where stars spend most of the residual lifetime.
With the black-dotted lines in panels (a) and (b), the mass-loss rate is fitted as a function of the stellar mass and effective temperature~(see Eqs. \ref{eq:mdot_MS} and \ref{eq:mdot_RSG}).}
\label{fig:mdot}
\end{center}
\end{figure*}

\begin{table}
\caption{The coefficients of the fitting functions for the mass-loss rates in the MS phase~(Eq. \ref{eq:mdot_MS}).}
\begin{center}
{\begin{tabular}{c|cccccc}
\hline
$Z/\zsun$ & $\alpha_1$ & $\alpha_2$ & $\beta_1$ & $\beta_2$ & $\gamma$ & $\log T_{\rm eff,min}$\\
\hline
0              & 1.6   & 4.03 & $10^4$ & 5.04  & 4  & 4.92 \\
$10^{-4}$ & 1.4   & 3.65 & 150    & 4.93 & 2 & 4.85 \\
$10^{-3}$ & 1.4   & 3.55 & 200    & 4.89 & 2  & 4.82 \\
$10^{-2}$ & 1.65 & 3.4   & 200    & 4.85 & 2  & 4.78 \\
$10^{-1}$ & 0.0   & 3.5   & 300    & 4.82 & 2  & 4.72 \\
\hline
\end{tabular}}
\end{center}
\label{tab:fit_mass_loss}
\end{table}

In Figure \ref{fig:mdot}, we show the pulsation-driven mass-loss rate as a function of the surface temperature for the same models with Figures \ref{fig:growth_rate_z2}~(panel a) and \ref{fig:growth_rate_m1e3}~(panel b).
In the MS stage, the mass-loss rate becomes the highest and decreases monotonically as the star evolves decreasing its surface temperature. This reflects the behavior of $|\sigma_{\rm I}|$ in Figures \ref{fig:growth_rate_z2} and \ref{fig:growth_rate_m1e3}~(see also Eq. \ref{eq:mdot_para}).
The mass-loss rate becomes higher with increasing metallicity and stellar mass, as expected from the dependence of $|\sigma_{\rm I}|$ and $E_{\rm puls}$ on these parameters.
For each metallicity case, the mass-loss rate can be fitted as a function of the stellar mass and effective temperature:
\begin{align}
\log \left(\frac{\dot{M}}{\msun\ {\rm yr}^{-1}}\right) =&~\alpha_1 \log \left(\frac{M_\ast}{10^3\ \msun}\right) - \alpha_2 \nonumber\\
& - \beta_1 \left(\log T_{\rm eff} - \beta_2 \right)^\gamma,
\label{eq:mdot_MS}
\end{align}
for $\log T_{\rm eff} \geq \log T_{\rm eff,min}$.
The coefficients, $\alpha_1$, $\alpha_2$, $\beta_1$, $\beta_2$, and $\gamma$, and $\log T_{\rm eff,min}$ are presented in Table \ref{tab:fit_mass_loss}.
Note that the metallicity dependence of $\alpha_2$ and $\beta_2$ can also be fitted as: $\alpha_2 = 3.65-0.125 \left(\log(Z/Z_\odot)+4.0\right)$ and $\beta_2 = 4.93-0.04 \left(\log(Z/Z_\odot)+4.0\right)$ in the metallicity range of $10^{-4} \lesssim Z/Z_\odot \lesssim 10^{-2}$.

In the RSG phase, the mass-loss rate increases monotonically as the envelope expands and the surface temperature decreases, and reaches $\gtrsim 10^{-3}~\msunyr$ in the later phase.
While the mass-loss rate becomes higher for more massive models~(panel a), it is almost independent of the metallicity~(panel b).
Namely, the mass-loss rate can be approximated as
\begin{align}
\log \left(\frac{\dot{M}}{\msun\ {\rm yr}^{-1}}\right) =&  -2.88 + \log \left(\frac{M_\ast}{10^3\ \msun}\right) \nonumber\\
& - 15.6\left(\log T_{\rm eff} - 3.7 \right).
\label{eq:mdot_RSG}
\end{align}
Note that the fitting formula is valid at $\log T_{\rm eff} \leq 3.85$~(and $3.7$) for $Z/\zsun \lesssim 10^{-2}$~(and $\simeq 10^{-1}$, respectively).

From the mass-loss formulae derived above, we can estimate how much fraction of the initial mass is lost during the evolution.
For example, in the case of $Z = 10^{-2}\ \zsun$, by assuming that the mass loss in the MS stage at $\log T_{\rm eff} = 4.85$ continues over the entire MS life of $\sim$ Myr and that the mass loss in the RSG stage at $\log T_{\rm eff} = 3.7$ lasts over the remaining life of $\sim$ 0.1 Myr, the fraction of mass lost can be estimated as $\Delta M/M_{\rm ZAMS} \sim 0.4 \left(M_{\rm ZAMS}/10^3\ \msun\right)^{0.65} + 0.13$, which is in the order of $\sim$ 30-100\% for $M_{\rm ZAMS}/\msun = 300\mbox{-}3000$.
Therefore, to reveal the final mass of massive stars, the feedback from pulsation-driven mass loss should be considered in the evolution calculations.

\vspace{5mm}
\subsection{Pulsation-driven mass loss throughout the evolution}\label{subsec:mass_loss}

\begin{figure*}
\begin{center}
\begin{tabular}{cc}
{\includegraphics[scale=1.1]{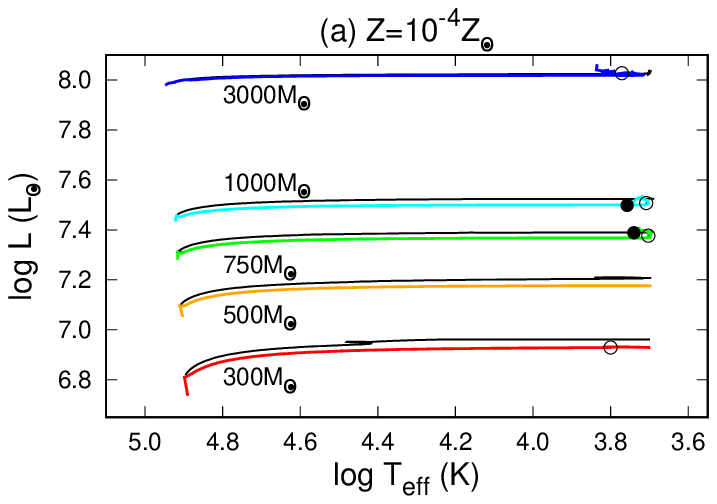}}
{\includegraphics[scale=1.1]{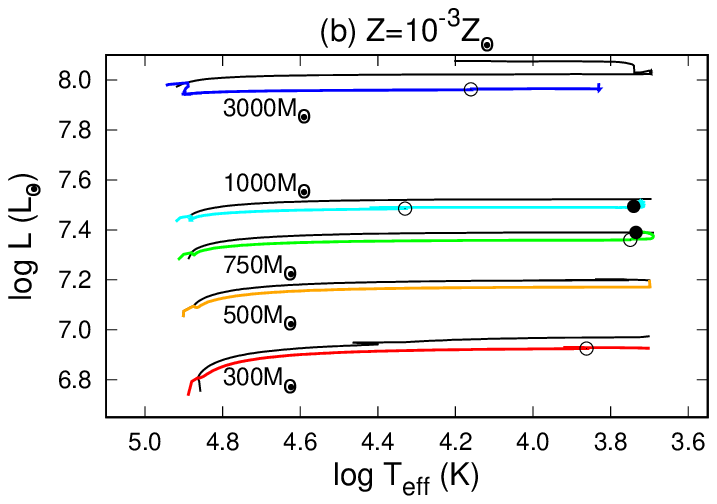}}\\
{\includegraphics[scale=1.1]{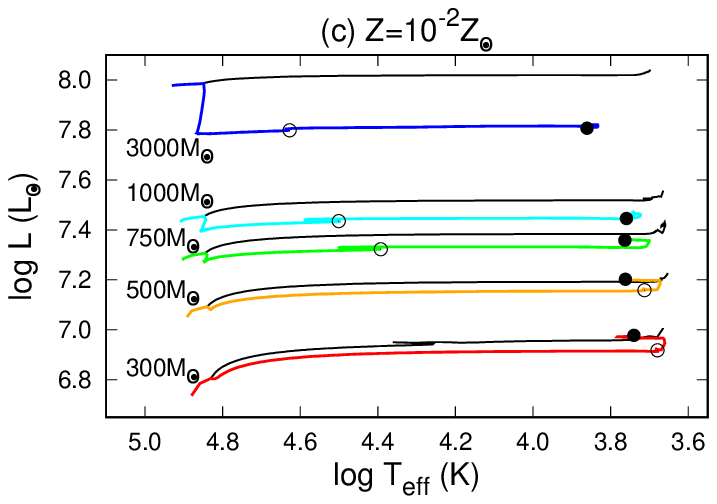}}
{\includegraphics[scale=1.1]{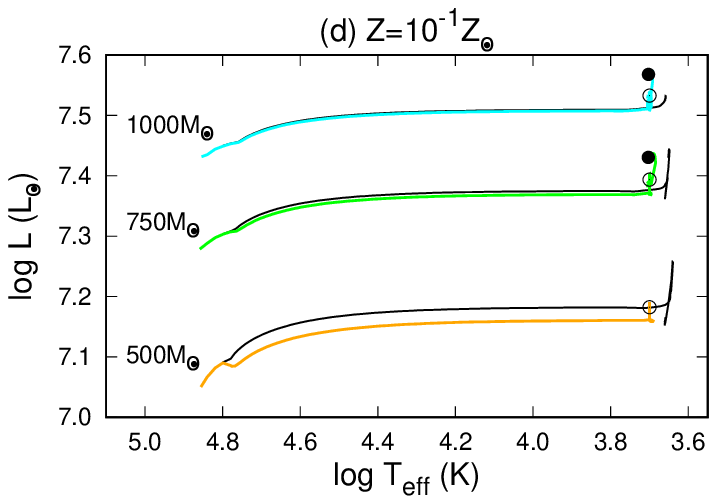}}\\
\end{tabular}
\caption{Evolutionary tracks in the HR diagram for the stellar models including feedback from the pulsation-driven wind~(colored curves).
In each panel, the results without mass loss are also shown by black curves, for comparison.
Panels (a)-(d) show the results for $Z/\zsun = 10^{-4}, 10^{-3}, 10^{-2}$, and $10^{-1}$, respectively.
Open and filled circles indicate the epochs of hydrogen and helium depletion in the core, respectively.
Pulsation-driven mass loss reduces the stellar mass and luminosity more significantly for higher mass cases.}
\label{fig:hrd_massloss}
\end{center}
\end{figure*}

Next, we calculate the stellar evolution models including pulsation-driven mass loss.
The mass-loss rate is given by the fitting formulae shown in Eqs. \eqref{eq:mdot_MS} and \eqref{eq:mdot_RSG}.
In Figure \ref{fig:hrd_massloss}, we show the evolutionary tracks of stellar models with pulsation-driven mass loss for different metallicities, $Z/\zsun =$ (a) $10^{-4}$, (b) $10^{-3}$, (c) $10^{-2}$, and (d) $10^{-1}$.
For comparison, the cases without mass loss are overlaid in the HR diagram~(black curves).
Open and filled circles indicate the epochs of hydrogen and helium depletion in the core, respectively.

In the early MS phase, all the stellar models reduce their masses by pulsation-driven winds, lowering their luminosities.
Due to the mass-loss process, stars keep their surface temperature higher and remain unstable for a longer time, compared to the models without mass loss.
During the expansion phase~(i.e., in the Hertzsprung gap), the stabilized stars evolve as less massive stars with lower luminosities.
In the RSG phase, the pulsation-driven wind sets in by the $\kappa$-mechanism and the mass-loss rate increases as the surface temperature decreases.
Once the surface temperature reaches $\log T_{\rm eff} \sim 3.7$, the mass loss becomes so strong that stars show blueward evolution to $\log T_{\rm eff} \gtrsim 3.7$ and the mass-loss rate is self-regulated at $\dot{M} \lesssim 10^{-3}~\msunyr$.

\begin{figure*}
\begin{center}
\begin{tabular}{cc}
{\includegraphics[scale=1.1]{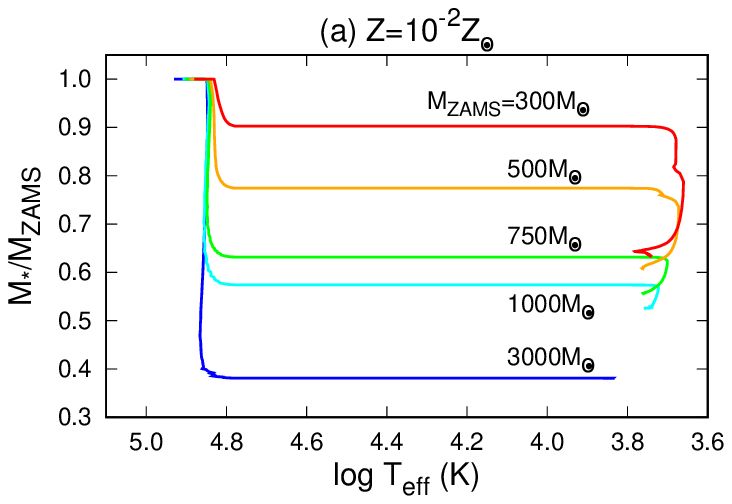}}
{\includegraphics[scale=1.1]{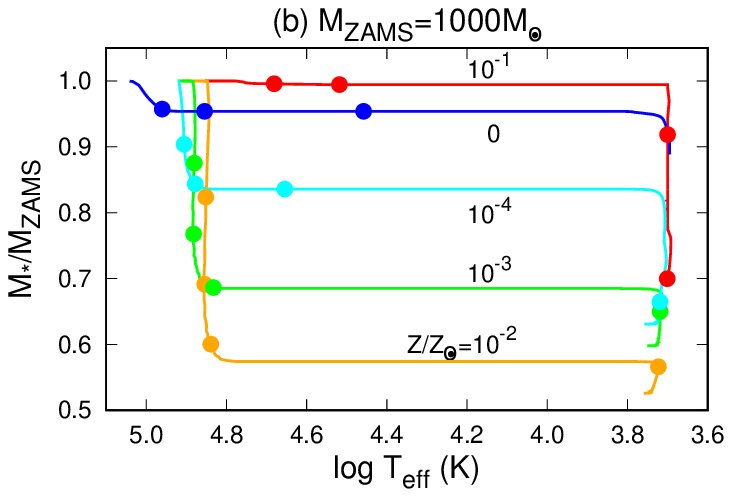}}
\end{tabular}
\caption{The time evolution of the stellar mass as a function of the surface temperature, for the models of $Z = 10^{-2}\ \zsun$ with various stellar masses~(panel a) and those of $M_{\rm ZAMS} = 1000\ \msun$ with various metallicities~(panel b).
In panel (b), filled circles on each track mark the evolutionary epochs in every 0.5 Myr starting from the ZAMS stage.
Overall, more massive and higher metallicity stars tend to maintain a smaller fraction of the initial mass at the end of lifetime.
Note that in panel (b), the $Z = 10^{-1}\ \zsun$ star does not suffer from mass loss in the MS stage but lose $\sim 30$\% of the mass in the RSG stage~(see text).}
\label{fig:total_loss}
\end{center}
\end{figure*}

In Figure \ref{fig:total_loss}, we show the evolution of the stellar mass normalized by the ZAMS mass as a function of the surface temperature for the models of $Z = 10^{-2}\ \zsun$ with various stellar masses~(panel a) and those of $M_{\rm ZAMS} = 1000\ \msun$ with various metallicities~(panel b).
In panel (b), evolutionary epochs in every 0.5 Myr starting from the ZAMS stage are shown on each track with filled circles.
We note that in Figure \ref{fig:total_loss}, the evolution is followed until helium core depletion for all the models except the $Z=0$ case, where the simulation is terminated at 1.8 Myr.

In the $Z=10^{-2}\ \zsun$ models~(panel a), during the MS, since the mass-loss rate increases with the stellar mass~(Figure \ref{fig:mdot}), more massive stars maintain a smaller fraction of the initial mass at the end of the MS.
On the other hand, their RSG stages last longer for lower masses and thus a larger fraction of the stellar mass is lost.
In total, $\sim 65\mbox{-}40\%$ of the initial mass is lost by helium exhaustion in the $M_{\rm ZAMS}/\msun = 300\mbox{-}3000$ models.
Even for the most massive case, a remnant BH as massive as $\gtrsim 1000\ \msun$ can be left.

In the $M_{\rm ZAMS} = 1000\ \msun$ models~(panel b), the low metallicity stars with $Z \leq 10^{-2}\ Z_\odot$ lose a larger fraction of the mass with increasing metallicity~($\sim 5\mbox{-}40$\% for $Z=0\mbox{-}10^{-2}\ \zsun$) and reside in the instability region for $\sim 1.5$ Myr, which is extended due to the mass loss from $\sim 0.5$ Myr in the cases without mass loss.
Their residual lifetime is not so long that the fraction of the mass lost in the RSG stage is subdominant or at most comparable to that in the MS stage.
On the other hand, the highest metallicity star~($Z=10^{-1}\ Z_\odot$) begins its evolution with the lowest surface temperature in the ZAMS stage and leaves the instability region of the MS stage significantly earlier~($< 0.5$ Myr) with negligible mass loss.
In the RSG stage, the star becomes unstable and loses $\sim 30$\% of the mass in the last $\gtrsim 0.5$ Myr of the lifetime.
In summary, for all the metallicities, $ > 50$\% of the ZAMS mass is left by the time of helium exhaustion, and thus the remnant BH is expected to be as massive as $\gtrsim 500\ \msun$.

\begin{figure}
\begin{center}
\includegraphics[scale=1.2]{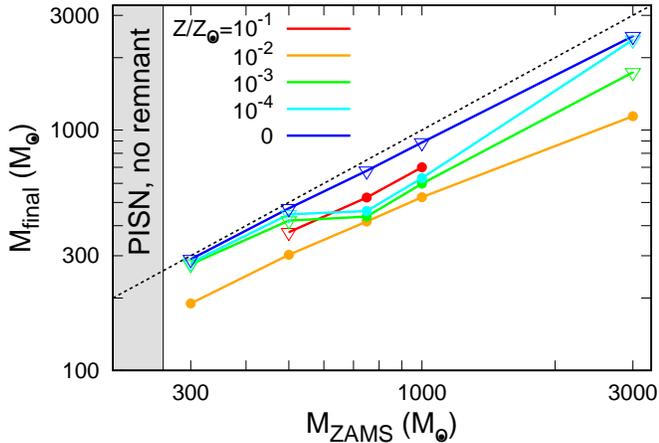}
\caption{The relation between the final stellar mass and the ZAMS mass.
Each colored curve corresponds to the different metallicities shown in the legend, and the black-dotted line to $M_{\rm final} = M_{\rm ZAMS}$.
Filled circles~(or open triangles) indicate the results for the models where the computations are~(or are not) performed until helium core depletion.
In the grey-shaded region, no remnants are left because of PISNe.}
\label{fig:zams_vs_final}
\end{center}
\end{figure}

In Figure \ref{fig:zams_vs_final}, we summarize the relation between the final and ZAMS mass for all the models where pulsation-driven mass-loss rate is taken into account self-consistently.
For the models where their simulations are followed until~(or terminated before) helium core depletion, the results are shown by the filled circles~(or open triangles, respectively).
When the ZAMS mass is less massive than $\sim 260~\msun$~(grey-shaded region), no remnants are left because of pair-instability supernovae~(PISN)~\citep{Woosley2002,Woosley2017}.
We find that pulsation-driven mass loss becomes the strongest in the cases of $Z=10^{-2}\ \zsun$~(orange).
In all the metallicity cases, the final stellar mass can be more massive than $M_{\rm final}/\msun \sim 200\mbox{-}1200$ for the stars with the ZAMS mass $M_{\rm ZAMS}/\msun = 300\mbox{-}3000$. 

\begin{figure*}
\begin{center}
\begin{tabular}{cc}
{\includegraphics[scale=1.1]{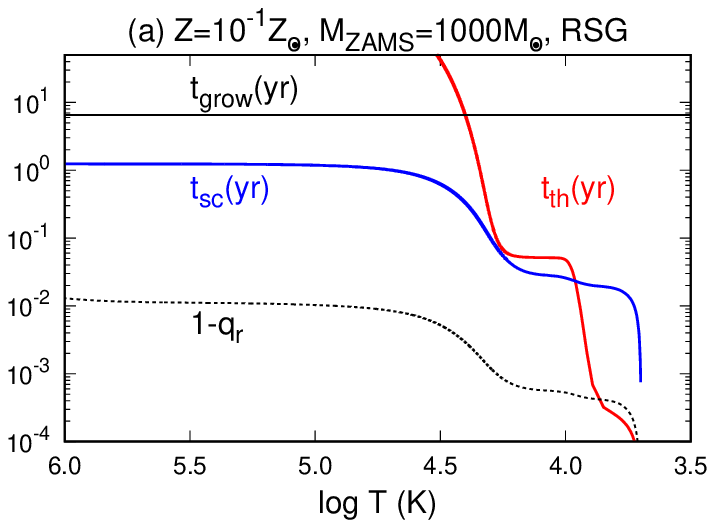}}
{\includegraphics[scale=1.1]{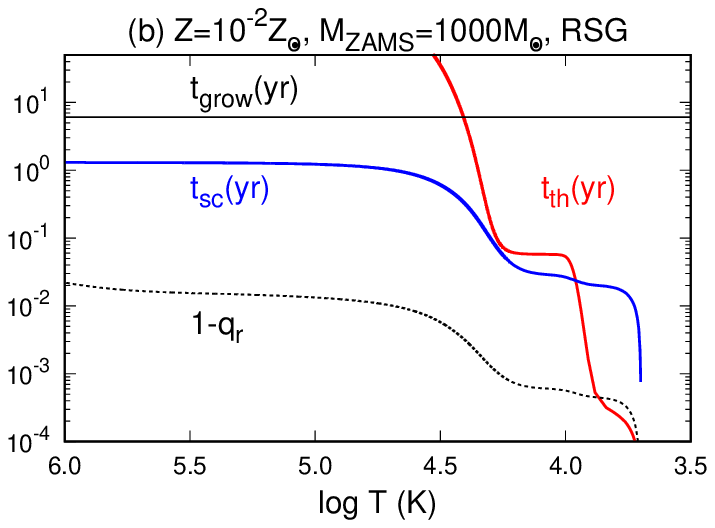}}
\end{tabular}
\caption{The radial profiles of the thermal timescale~($t_{\rm th}$, red solid), sound-crossing time~($t_{\rm sc}$, blue solid), and the fraction of the stellar mass contained outside a radius $r$~($1-q_r$; black dashed), as a function of temperature for the RSG models with $M_{\rm ZAMS} = 1000\ \msun$.
In panels (a) and (b), the cases with $Z/Z_\odot = 10^{-2}$ and $10^{-1}$ are shown, respectively.
Both the thermal and sound-crossing timescales in the instability-driving layer~($\log T \sim 4.0$) are substantially shorter than the instability growth timescale~($t_{\rm grow} \sim 10$ yr, black solid).
Thus, a new equilibrium state can be quickly established before the subsequent mass eruption occurs.}
\label{fig:timescale}
\end{center}
\end{figure*}

It is worth studying how quickly the structure of a pulsating star settles down to a dynamically and thermally relaxed configuration.
In Figure \ref{fig:timescale}, we show the radial profiles of the thermal timescale ($t_{\rm th}$), sound crossing time ($t_{\rm sc}$), and the fraction of the stellar mass contained outside a radius $r$, $(1-q_r)$, as a function of temperature.
Here, those two timescales are defined as
\begin{equation}
t_{\rm th}(r) = \int_r^{R_\ast} 4 \pi c_P T \rho r^2 dr/L_{\rm rad}
\label{eq:t_th}
\end{equation}
and
\begin{equation}
t_{\rm sc}(r) = \int_r^{R_\ast} dr/c_{\rm s}.
\label{eq:t_sound}
\end{equation}
For the RSG models, the mass injected into pulsation-driven winds in each eruption event~($t_{\rm grow} \sim 10$ yrs) is as high as $\sim 0.03\ \msun$, which is contained in the outer-most layer~($\log T<3.8$) but above the instability-driving layer~($\log T \sim 4.0$; see also Figure \ref{fig:work_integral}).
Since both the thermal~(red) and sound-crossing~(blue) timescales in the layer are much shorter than the instability growth timescale~(black solid; $t_{\rm grow} \sim 10$ yr), a new equilibrium state can be established before the subsequent mass eruption occurs.

\vspace{10mm}
\section{Summary and Discussion}\label{sec:summary}

Very massive stars with $M_{\rm ZAMS} \sim 10^2\mbox{-}10^4\ \msun$ formed via runaway collisions in dense star clusters have attracted attention as possible progenitors of massive seeds for high-$z$ SMBHs.
However, whether or not the massive merger products can keep their masses within their stellar lifetime without significant mass loss is poorly understood.
Here, we study the pulsational stability of very massive stars with a wide range of the ZAMS mass at $300 \leq M_{\rm ZAMS}/\msun \leq 3000$ and stellar metallicity $0\leq Z/\zsun \leq 10^{-1}$, which are relevant to the stellar runway merger scenario.
Conducting the stability analysis to stellar structure models obtained with the MESA code, we estimate the masses of merger remnants left at the centers of dense clusters.
Our findings are summarized below:

\begin{itemize}

\item All the stellar ZAMS models are pulsationally unstable, owing to the $\epsilon$-mechanism driven by nuclear burning in the cores~(Figures \ref{fig:hrd_unstable} and \ref{fig:work_integral}).
As the stars evolve off the ZAMS, they are stabilized because of radiative damping in the envelope.
The pulsational instability grows in $\sim 10^3\ {\rm yrs}$, which is significantly shorter than the lifetime of massive stars~(Figures \ref{fig:growth_rate_z2} and \ref{fig:growth_rate_m1e3}).
The mass-loss rate is estimated in the range of $\sim 10^{-6}\mbox{-}10^{-3}\ \msun\ {\rm yr}^{-1}$~(Figure \ref{fig:mdot}, Eq. \ref{eq:mdot_MS}, and Table \ref{tab:fit_mass_loss}).
Both the growth rate and mass-loss rate increase with stellar mass and metallicity.

\item In the RSG stages, all the stellar models are destabilized again, owing to the $\kappa$-mechanism driven in the ionization zones of hydrogen~(Figures \ref{fig:hrd_unstable} and \ref{fig:work_integral}).
The instability grows in a much shorter timescale of $\sim 10\ {\rm yrs}$ compared to that in the MS phase~(Figures \ref{fig:growth_rate_z2} and \ref{fig:growth_rate_m1e3}).
The mass-loss rate in the RSG phase rises to $\sim 10^{-3}\mbox{-}10^{-2}\ \msun\ {\rm yr}^{-1}$ as the surface temperature is lowered~(Figure \ref{fig:mdot}).
For a fixed stellar mass, both the growth rate and mass-loss rate are almost independent of the metallicity.
The mass-loss rate is well approximated by Eq. \eqref{eq:mdot_RSG}.

\item Adopting the mass-loss rate obtained from the linear stability analysis, we recalculate the stellar structure evolution (Figure \ref{fig:hrd_massloss}).
For the models with $M_{\rm ZAMS} = 1000\ \msun$ and various metallicities, the total amount of mass loss becomes the largest in the case of $Z=10^{-2}\ \zsun$~(Figures \ref{fig:total_loss} and \ref{fig:zams_vs_final}).
Even in this case, the final mass is more than $\sim$ 50\% of the initial ZAMS mass, and thus the remnant BH is expected to be as massive as $\gtrsim 500\ \msun$.
For the models with $Z=10^{-2}\ \zsun$ and different initial masses, the total amount of mass loss increases with the stellar mass.
While the most massive case of $M_{\rm ZAMS} = 3000\ \msun$ loses $\sim$ 60\% of the ZAMS mass, the remnant BH can still be as massive as $\gtrsim 1000\ \msun$.

\end{itemize}

\vspace{5mm}

\begin{figure}
\begin{center}
\includegraphics[scale=1.1]{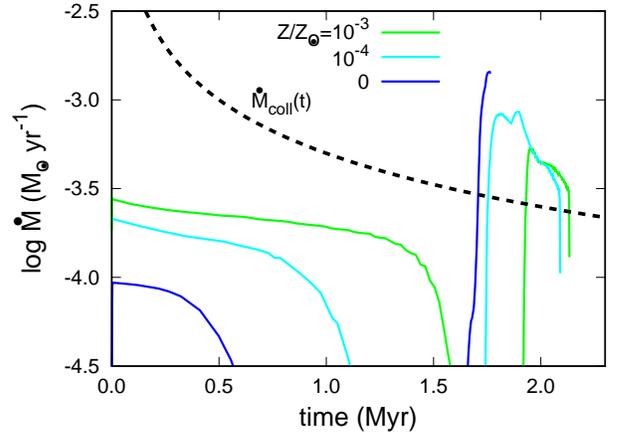}
\caption{The comparison of the mass-loss rate for the $M_{\rm ZAMS} = 1000\ \msun$ models~(solid curves) with the averaged mass-gain rate $\dot{M}_{\rm coll}$ via runaway stellar collisions in a star cluster~(dashed curve).
Individual colors indicate the cases of $Z/\zsun = 0$~(blue), $10^{-4}$~(light blue), and $10^{-3}$~(green), respectively.
The mass-gain rate exceeds the mass-loss rate in the MS stage~($t<1.6$ Myr), during which the central star grows up to $\sim 1500\ \msun$.
In the RSG stage~($t \gtrsim 1.6$ Myr), while the mass loss overcomes the mass accretion, reducing the stellar mass by a few $100\ \msun$, the final stellar mass is still as massive as $\gtrsim 1000\ \msun$.}
\label{fig:gain_vs_loss}
\end{center}
\end{figure}

Here, we discuss how pulsation-driven mass loss affects the evolution of massive stars via runaway stellar mergers in dense star clusters.
Although star cluster formation in the early universe is still very uncertain, previous authors presumed very metal-poor environments of $Z/\zsun \sim 10^{-5}\mbox{-}10^{-3}$ as the cluster formation sites~\citep{Omukai2008,Devecchi2009,Katz2015,Sakurai2017}.
A series of $N$-body simulations by \cite{Sakurai2017} show that in a cluster of $\sim 10^5\ \msun$ formed in a young protogalaxy, 
frequent stellar collisions with an interval of $\sim 0.1\ {\rm Myr}$ lead to growth of the most massive object at the center at a rate of
\begin{equation}
\dot{M}_{\rm coll}(t) = 4 \times 10^{-4} \left(\frac{M_{\rm cl}}{10^5\ \msun}\right) \left(\frac{t}{1\ {\rm Myr}}\right)^{-1}\ \msun\ {\rm yr}^{-1},
\label{eq:mass_growth}
\end{equation}
where $M_{\rm cl}$ is the cluster mass, and $t$ is the time from the cluster formation \citep[see also][]{Portegies_Zwart2002}.
While the stellar envelope of the merger product would be bloated like the structure of a RSG, the star contracts quickly to a thermally-relaxed MS structure in a Kelvin-Helmholtz time of $< 10^4$ yrs and turns pulsationally unstable in $\sim 10^3$ yrs.
Therefore, our stability analyses against thermally-relaxed stellar structures are justified, even in the runaway stellar-merger scenario.

In Figure \ref{fig:gain_vs_loss}, we show the mass-loss rates due to stellar pulsation~(solid lines) for the $M_{\rm ZAMS} = 1000\ \msun$ models with $Z/\zsun = 0$~(blue), $10^{-4}$~(light blue), and $10^{-3}$~(green), respectively.
As a reference, the mass-gain rate~(Eq. \ref{eq:mass_growth}; dashed line) is also overlaid.
In the MS phase~($t<1.6$ Myr), the mass-gain rate exceeds the mass-loss rate for all the metallicity cases.
During this period, the central star grows up to $\sim 1500\ \msun$~\citep{Sakurai2017}, according to Eq. \eqref{eq:mass_growth}.
In the late RSG phase ($t \gtrsim 1.6$ Myr), however, the star begins to lose mass via stellar pulsation significantly, overcoming the accretion rate via stellar mergers.
Our stability analysis suggests that a RSG star with $\sim 1000\ \msun$ becomes unstable and lose $\sim 10\mbox{-}20\%$ of its mass in the cases of $Z/\zsun \lesssim 10^{-3}$~(Figure \ref{fig:total_loss}a).
Therefore, we conclude that the star can leave behind a remnant BH as massive as $\gtrsim 1000\ \msun$, which would grow to an SMBH with $\sim 10^9\ \msun$ via the Eddington-limited accretion within a Hubble timescale at $z>6$.
Note that to estimate the final remnant mass more quantitatively, the evolution of merging stars should be calculated self-consistently by taking account of the mass loss associated with stellar mergers~\citep{Glebbeek2009}.

\ 

The mass-loss rate estimated from Eq. \eqref{eq:mdot_puls} is based on the linear stability analysis, while mass loss occurs after the perturbation grows to a non-linear regime, where strong shock waves would make the stellar surface structure deviate from the unperturbed structure significantly.
Non-linear hydrodynamical simulations of stellar pulsation are required to derive a reliable mass-loss prescription~\citep[e.g.,][]{Yadav2018}.
Moreover, the estimate of mass-loss rates with Eq. \eqref{eq:mdot_puls} was proposed by 
\cite{Appenzeller1970a,Appenzeller1970b} for massive stars with $100\lesssim M_{\rm ZAMS}/\msun \lesssim 600$ 
and $Z=1.5~\zsun$.
To explore the mass-loss prescription for lower-metallicity stars is left for future investigations.

Our stability analysis has been carried out by adopting the frozen-in approximation where the interaction between pulsation and convective motions is neglected.
This approximation becomes invalid in the MS cores where nuclear burning is activated and in the deep convective envelopes of RSGs with $\log T_{\rm eff} < 3.7$.
\cite{Shiode2012} found that pulsation damping in convective zones overcomes the excitation by the $\epsilon$-mechanism, while it is not so strong as to suppress unstable pulsation.
However, those results depend on uncertain model parameters to treat time-dependent convective energy transport
\citep[e.g.,][]{Houdek2015}.
Since convective motions are intrinsically multi-dimensional, its effect should be studied in more detail by multidimensional radiation hydrodynamical calculations, such as performed in the asymptotic giant branch stars~\citep[e.g.,][]{Freytag2017}.

So far, we have focused on the effect of pulsation-driven mass loss on the stellar evolution.
For metal-enriched massive stars, the acceleration of winds by the line force due to bound-bound absorption may not be neglected~\citep[e.g.,][]{Castor1975,Bowen1988}.
For a MS star with $\log T_{\rm eff} \gtrsim 4.0$, the line-driven mass-loss rate is theoretically derived by \cite{Vink2001}.
The mass-loss rate has the maximum value of $\sim 8.3 \times 10^{-6} \left(M_{\rm ZAMS}/10^3\ \msun\right)^{0.88} \left(Z/10^{-2}\ \zsun\right)^{0.85}\ \msun\ {\rm yr}^{-1}$ at $\log T_{\rm eff} \sim 4.6$.
Even if the star suffers from the wind over the entire MS lifetime, the fraction of the ejected mass is less than $\sim 1\%$ of the original ZAMS mass.
For a RSG star with $\log T_{\rm eff} \lesssim 4.0$, while the driving mechanism is still highly uncertain, the mass-loss formula is empirically obtained by \cite{Nieuwenhuijzen1990}.
The rate increases to $\dot{M}\simeq 3.7 \times 10^{-3} \left(M_{\rm ZAMS}/10^3\ \msun\right)^{1.8} \left(Z/10^{-2}\ \zsun\right)^{0.85}\ \msun\ {\rm yr}^{-1}$ as the surface temperature drops to $\log T_{\rm eff} \sim 3.7$.
If the stellar mass is continuously lost over the entire RSG stage with a shorter duration of $\sim 0.1$ Myr, the fraction of the ejected mass is $\Delta M_{\rm RSG}/M_{\rm ZAMS} \sim 0.37 \left(M_{\rm ZAMS}/10^3\ \msun\right)^{0.8} \left(Z/10^{-2}\ \zsun\right)^{0.85}$.
Therefore, for lower metallicity cases $Z \lesssim 10^{-3}~\zsun$, pulsation-driven mass loss would be a dominant process to determine the final mass of a massive merger product.

\vspace{30mm}
\section*{acknowledgments}
The authors wish to express their cordial thanks to Profs. Hideyuki Saio, Zoltan Haiman, and Takashi Yoshida for their constructive suggestions and comments.
We also thank Ryosuke Hirai and Hiroto Mitani for the instruction of the MESA code.
Numerical calculations are carried out with the computer cluster, {\tt Draco}, supported by the Frontier Research Institute for Interdisciplinary Sciences in Tohoku University, and with High-performance Computing Platform of Peking University.
This work is supported in part by MEXT/JSPS KAKENHI grants~(17H01102, 17H02869, 17H06360: KO), the National Science Foundation of China (11721303, 11991052, 11950410493; KI), and the National Key R\&D Program of China (2016YFA0400702; KI).


%




\bibliographystyle{aasjournal}

\begin{thebibliography}{}
\expandafter\ifx\csname natexlab\endcsname\relax\def\natexlab#1{#1}\fi
\providecommand{\url}[1]{\href{#1}{#1}}
\providecommand{\dodoi}[1]{doi:~\href{http://doi.org/#1}{\nolinkurl{#1}}}
\providecommand{\doeprint}[1]{\href{http://ascl.net/#1}{\nolinkurl{http://ascl.net/#1}}}
\providecommand{\doarXiv}[1]{\href{https://arxiv.org/abs/#1}{\nolinkurl{https://arxiv.org/abs/#1}}}

\bibitem[{{Agarwal} {et~al.}(2012){Agarwal}, {Khochfar}, {Johnson}, {Neistein},
  {Dalla Vecchia}, \& {Livio}}]{Agarwal2012}
{Agarwal}, B., {Khochfar}, S., {Johnson}, J.~L., {et~al.} 2012, MNRAS, 425,
  2854, \dodoi{10.1111/j.1365-2966.2012.21651.x}

\bibitem[{{Alister Seguel} {et~al.}(2020){Alister Seguel}, {Schleicher},
  {Boekholt}, {Fellhauer}, \& {Klessen}}]{Seguel2020}
{Alister Seguel}, P.~J., {Schleicher}, D.~R.~G., {Boekholt}, T.~C.~N.,
  {Fellhauer}, M., \& {Klessen}, R.~S. 2020, MNRAS, 493, 2352,
  \dodoi{10.1093/mnras/staa456}

\bibitem[{{Alvarez} {et~al.}(2009){Alvarez}, {Wise}, \& {Abel}}]{Alvarez2009}
{Alvarez}, M.~A., {Wise}, J.~H., \& {Abel}, T. 2009, ApJ, 701, L133,
  \dodoi{10.1088/0004-637X/701/2/L133}

\bibitem[{{Appenzeller}(1970{\natexlab{a}})}]{Appenzeller1970a}
{Appenzeller}, I. 1970{\natexlab{a}}, A\&A, 5, 355

\bibitem[{{Appenzeller}(1970{\natexlab{b}})}]{Appenzeller1970b}
---. 1970{\natexlab{b}}, A\&A, 9, 216

\bibitem[{{Ba{\~n}ados} {et~al.}(2018){Ba{\~n}ados}, {Venemans},
  {Mazzucchelli}, {Farina}, {Walter}, {Wang}, {Decarli}, {Stern}, {Fan},
  {Davies}, {Hennawi}, {Simcoe}, {Turner}, {Rix}, {Yang}, {Kelson}, {Rudie}, \&
  {Winters}}]{Banados2018}
{Ba{\~n}ados}, E., {Venemans}, B.~P., {Mazzucchelli}, C., {et~al.} 2018,
  Nature, 553, 473, \dodoi{10.1038/nature25180}

\bibitem[{{Baraffe} {et~al.}(2001){Baraffe}, {Heger}, \&
  {Woosley}}]{Baraffe2001}
{Baraffe}, I., {Heger}, A., \& {Woosley}, S.~E. 2001, ApJ, 550, 890,
  \dodoi{10.1086/319808}

\bibitem[{{Becerra} {et~al.}(2015){Becerra}, {Greif}, {Springel}, \&
  {Hernquist}}]{Becerra2015}
{Becerra}, F., {Greif}, T.~H., {Springel}, V., \& {Hernquist}, L.~E. 2015,
  MNRAS, 446, 2380, \dodoi{10.1093/mnras/stu2284}

\bibitem[{{Boekholt} {et~al.}(2018){Boekholt}, {Schleicher}, {Fellhauer},
  {Klessen}, {Reinoso}, {Stutz}, \& {Haemmerl{\'e}}}]{Boekholt2018}
{Boekholt}, T.~C.~N., {Schleicher}, D.~R.~G., {Fellhauer}, M., {et~al.} 2018,
  MNRAS, 476, 366, \dodoi{10.1093/mnras/sty208}

\bibitem[{{Bowen}(1988)}]{Bowen1988}
{Bowen}, G.~H. 1988, ApJ, 329, 299, \dodoi{10.1086/166378}

\bibitem[{{Bromm} \& {Loeb}(2003)}]{Bromm_Loeb2003}
{Bromm}, V., \& {Loeb}, A. 2003, ApJ, 596, 34, \dodoi{10.1086/377529}

\bibitem[{{Castor} {et~al.}(1975){Castor}, {Abbott}, \& {Klein}}]{Castor1975}
{Castor}, J.~I., {Abbott}, D.~C., \& {Klein}, R.~I. 1975, ApJ, 195, 157,
  \dodoi{10.1086/153315}

\bibitem[{{Chon} \& {Omukai}(2020)}]{Chon2020}
{Chon}, S., \& {Omukai}, K. 2020, MNRAS, 494, 2851,
  \dodoi{10.1093/mnras/staa863}

\bibitem[{{Devecchi} \& {Volonteri}(2009)}]{Devecchi2009}
{Devecchi}, B., \& {Volonteri}, M. 2009, ApJ, 694, 302,
  \dodoi{10.1088/0004-637X/694/1/302}

\bibitem[{{Fan}(2006)}]{Fan2006}
{Fan}, X. 2006, NewAR, 50, 665, \dodoi{10.1016/j.newar.2006.06.077}

\bibitem[{{Freytag} {et~al.}(2017){Freytag}, {Liljegren}, \&
  {H{\"o}fner}}]{Freytag2017}
{Freytag}, B., {Liljegren}, S., \& {H{\"o}fner}, S. 2017, A\&A, 600, A137,
  \dodoi{10.1051/0004-6361/201629594}

\bibitem[{{Glebbeek} {et~al.}(2009){Glebbeek}, {Gaburov}, {de Mink}, {Pols}, \&
  {Portegies Zwart}}]{Glebbeek2009}
{Glebbeek}, E., {Gaburov}, E., {de Mink}, S.~E., {Pols}, O.~R., \& {Portegies
  Zwart}, S.~F. 2009, A\&A, 497, 255, \dodoi{10.1051/0004-6361/200810425}

\bibitem[{{Haemmerl{\'e}} {et~al.}(2018){Haemmerl{\'e}}, {Woods}, {Klessen},
  {Heger}, \& {Whalen}}]{Haemmerle2018}
{Haemmerl{\'e}}, L., {Woods}, T.~E., {Klessen}, R.~S., {Heger}, A., \&
  {Whalen}, D.~J. 2018, MNRAS, 474, 2757, \dodoi{10.1093/mnras/stx2919}

\bibitem[{{Heger} {et~al.}(1997){Heger}, {Jeannin}, {Langer}, \&
  {Baraffe}}]{Heger1997}
{Heger}, A., {Jeannin}, L., {Langer}, N., \& {Baraffe}, I. 1997, A\&A, 327,
  224.
\newblock \doarXiv{astro-ph/9705097}

\bibitem[{{Hirano} {et~al.}(2014){Hirano}, {Hosokawa}, {Yoshida}, {Umeda},
  {Omukai}, {Chiaki}, \& {Yorke}}]{Hirano2014}
{Hirano}, S., {Hosokawa}, T., {Yoshida}, N., {et~al.} 2014, ApJ, 781, 60,
  \dodoi{10.1088/0004-637X/781/2/60}

\bibitem[{{Hosokawa} {et~al.}(2016){Hosokawa}, {Hirano}, {Kuiper}, {Yorke},
  {Omukai}, \& {Yoshida}}]{Hosokawa2016}
{Hosokawa}, T., {Hirano}, S., {Kuiper}, R., {et~al.} 2016, ApJ, 824, 119,
  \dodoi{10.3847/0004-637X/824/2/119}

\bibitem[{{Hosokawa} {et~al.}(2011){Hosokawa}, {Omukai}, {Yoshida}, \&
  {Yorke}}]{Hosokawa2011}
{Hosokawa}, T., {Omukai}, K., {Yoshida}, N., \& {Yorke}, H.~W. 2011, Science,
  334, 1250, \dodoi{10.1126/science.1207433}

\bibitem[{{Hosokawa} {et~al.}(2013){Hosokawa}, {Yorke}, {Inayoshi}, {Omukai},
  \& {Yoshida}}]{Hosokawa2013}
{Hosokawa}, T., {Yorke}, H.~W., {Inayoshi}, K., {Omukai}, K., \& {Yoshida}, N.
  2013, ApJ, 778, 178, \dodoi{10.1088/0004-637X/778/2/178}

\bibitem[{{Houdek} \& {Dupret}(2015)}]{Houdek2015}
{Houdek}, G., \& {Dupret}, M.-A. 2015, Living Reviews in Solar Physics, 12, 8,
  \dodoi{10.1007/lrsp-2015-8}

\bibitem[{{Inayoshi} \& {Haiman}(2014)}]{Inayoshi_Haiman2014}
{Inayoshi}, K., \& {Haiman}, Z. 2014, MNRAS, 445, 1549,
  \dodoi{10.1093/mnras/stu1870}

\bibitem[{{Inayoshi} {et~al.}(2013){Inayoshi}, {Hosokawa}, \&
  {Omukai}}]{Inayoshi2013}
{Inayoshi}, K., {Hosokawa}, T., \& {Omukai}, K. 2013, MNRAS, 431, 3036,
  \dodoi{10.1093/mnras/stt362}

\bibitem[{{Inayoshi} \& {Omukai}(2012)}]{Inayoshi2012}
{Inayoshi}, K., \& {Omukai}, K. 2012, MNRAS, 422, 2539,
  \dodoi{10.1111/j.1365-2966.2012.20812.x}

\bibitem[{{Inayoshi} {et~al.}(2014){Inayoshi}, {Omukai}, \&
  {Tasker}}]{Inayoshi_Omukai2014}
{Inayoshi}, K., {Omukai}, K., \& {Tasker}, E. 2014, MNRAS, 445, L109,
  \dodoi{10.1093/mnrasl/slu151}

\bibitem[{{Inayoshi} {et~al.}(2019){Inayoshi}, {Visbal}, \&
  {Haiman}}]{Inayoshi2020}
{Inayoshi}, K., {Visbal}, E., \& {Haiman}, Z. 2019, arXiv e-prints,
  arXiv:1911.05791.
\newblock \doarXiv{1911.05791}

\bibitem[{{Katz} {et~al.}(2015){Katz}, {Sijacki}, \& {Haehnelt}}]{Katz2015}
{Katz}, H., {Sijacki}, D., \& {Haehnelt}, M.~G. 2015, MNRAS, 451, 2352,
  \dodoi{10.1093/mnras/stv1048}

\bibitem[{{Kippenhahn} {et~al.}(2012){Kippenhahn}, {Weigert}, \&
  {Weiss}}]{Kippenhahn2012}
{Kippenhahn}, R., {Weigert}, A., \& {Weiss}, A. 2012, {Stellar Structure and
  Evolution}, \dodoi{10.1007/978-3-642-30304-3}

\bibitem[{{Langer} {et~al.}(1983){Langer}, {Fricke}, \&
  {Sugimoto}}]{Langer1983}
{Langer}, N., {Fricke}, K.~J., \& {Sugimoto}, D. 1983, A\&A, 126, 207

\bibitem[{{Matsuoka} {et~al.}(2018){Matsuoka}, {Onoue}, {Kashikawa}, {Iwasawa},
  {Strauss}, {Nagao}, {Imanishi}, {Lee}, {Akiyama}, {Asami}, {Bosch},
  {Foucaud}, {Furusawa}, {Goto}, {Gunn}, {Harikane}, {Ikeda}, {Izumi},
  {Kawaguchi}, {Kikuta}, {Kohno}, {Komiyama}, {Lupton}, {Minezaki}, {Miyazaki},
  {Morokuma}, {Murayama}, {Niida}, {Nishizawa}, {Oguri}, {Ono}, {Ouchi},
  {Price}, {Sameshima}, {Schulze}, {Shirakata}, {Silverman}, {Sugiyama},
  {Tait}, {Takada}, {Takata}, {Tanaka}, {Tang}, {Toba}, {Utsumi}, \&
  {Wang}}]{Matsuoka2018}
{Matsuoka}, Y., {Onoue}, M., {Kashikawa}, N., {et~al.} 2018, PASJ, 70, S35,
  \dodoi{10.1093/pasj/psx046}

\bibitem[{{Milosavljevi{\'c}} {et~al.}(2009){Milosavljevi{\'c}}, {Bromm},
  {Couch}, \& {Oh}}]{Milos2009}
{Milosavljevi{\'c}}, M., {Bromm}, V., {Couch}, S.~M., \& {Oh}, S.~P. 2009, ApJ,
  698, 766, \dodoi{10.1088/0004-637X/698/1/766}

\bibitem[{{Montero} {et~al.}(2012){Montero}, {Janka}, \&
  {M{\"u}ller}}]{Montero2012}
{Montero}, P.~J., {Janka}, H.-T., \& {M{\"u}ller}, E. 2012, ApJ, 749, 37,
  \dodoi{10.1088/0004-637X/749/1/37}

\bibitem[{{Moriya} \& {Langer}(2015)}]{Moriya2015}
{Moriya}, T.~J., \& {Langer}, N. 2015, A\&A, 573, A18,
  \dodoi{10.1051/0004-6361/201424957}

\bibitem[{{Mortlock} {et~al.}(2011){Mortlock}, {Warren}, {Venemans}, {Patel},
  {Hewett}, {McMahon}, {Simpson}, {Theuns}, {Gonz{\'a}les-Solares}, {Adamson},
  {Dye}, {Hambly}, {Hirst}, {Irwin}, {Kuiper}, {Lawrence}, \&
  {R{\"o}ttgering}}]{Mortlock2011}
{Mortlock}, D.~J., {Warren}, S.~J., {Venemans}, B.~P., {et~al.} 2011, Nature,
  474, 616, \dodoi{10.1038/nature10159}

\bibitem[{{Nakauchi} {et~al.}(2017){Nakauchi}, {Hosokawa}, {Omukai}, {Saio}, \&
  {Nomoto}}]{Nakauchi2017}
{Nakauchi}, D., {Hosokawa}, T., {Omukai}, K., {Saio}, H., \& {Nomoto}, K. 2017,
  MNRAS, 465, 5016, \dodoi{10.1093/mnras/stw3114}

\bibitem[{{Nieuwenhuijzen} \& {de Jager}(1990)}]{Nieuwenhuijzen1990}
{Nieuwenhuijzen}, H., \& {de Jager}, C. 1990, A\&A, 231, 134

\bibitem[{{Oh} \& {Haiman}(2002)}]{Oh_Haiman2002}
{Oh}, S.~P., \& {Haiman}, Z. 2002, ApJ, 569, 558, \dodoi{10.1086/339393}

\bibitem[{{Omukai}(2001)}]{Omukai2001}
{Omukai}, K. 2001, ApJ, 546, 635, \dodoi{10.1086/318296}

\bibitem[{{Omukai} {et~al.}(2008){Omukai}, {Schneider}, \&
  {Haiman}}]{Omukai2008}
{Omukai}, K., {Schneider}, R., \& {Haiman}, Z. 2008, ApJ, 686, 801,
  \dodoi{10.1086/591636}

\bibitem[{{Onoue} {et~al.}(2019){Onoue}, {Kashikawa}, {Matsuoka}, {Kato},
  {Izumi}, {Nagao}, {Strauss}, {Harikane}, {Imanishi}, {Ito}, {Iwasawa},
  {Kawaguchi}, {Lee}, {Noboriguchi}, {Suh}, {Tanaka}, \& {Toba}}]{Onoue2019}
{Onoue}, M., {Kashikawa}, N., {Matsuoka}, Y., {et~al.} 2019, ApJ, 880, 77,
  \dodoi{10.3847/1538-4357/ab29e9}

\bibitem[{{Paxton} {et~al.}(2011){Paxton}, {Bildsten}, {Dotter}, {Herwig},
  {Lesaffre}, \& {Timmes}}]{Paxton2011}
{Paxton}, B., {Bildsten}, L., {Dotter}, A., {et~al.} 2011, ApJS, 192, 3,
  \dodoi{10.1088/0067-0049/192/1/3}

\bibitem[{{Paxton} {et~al.}(2013){Paxton}, {Cantiello}, {Arras}, {Bildsten},
  {Brown}, {Dotter}, {Mankovich}, {Montgomery}, {Stello}, {Timmes}, \&
  {Townsend}}]{Paxton2013}
{Paxton}, B., {Cantiello}, M., {Arras}, P., {et~al.} 2013, ApJS, 208, 4,
  \dodoi{10.1088/0067-0049/208/1/4}

\bibitem[{{Paxton} {et~al.}(2015){Paxton}, {Marchant}, {Schwab}, {Bauer},
  {Bildsten}, {Cantiello}, {Dessart}, {Farmer}, {Hu}, {Langer}, {Townsend},
  {Townsley}, \& {Timmes}}]{Paxton2015}
{Paxton}, B., {Marchant}, P., {Schwab}, J., {et~al.} 2015, ApJS, 220, 15,
  \dodoi{10.1088/0067-0049/220/1/15}

\bibitem[{{Paxton} {et~al.}(2018){Paxton}, {Schwab}, {Bauer}, {Bildsten},
  {Blinnikov}, {Duffell}, {Farmer}, {Goldberg}, {Marchant}, {Sorokina},
  {Thoul}, {Townsend}, \& {Timmes}}]{Paxton2018}
{Paxton}, B., {Schwab}, J., {Bauer}, E.~B., {et~al.} 2018, ApJS, 234, 34,
  \dodoi{10.3847/1538-4365/aaa5a8}

\bibitem[{{Paxton} {et~al.}(2019){Paxton}, {Smolec}, {Schwab}, {Gautschy},
  {Bildsten}, {Cantiello}, {Dotter}, {Farmer}, {Goldberg}, {Jermyn}, {Kanbur},
  {Marchant}, {Thoul}, {Townsend}, {Wolf}, {Zhang}, \& {Timmes}}]{Paxton2019}
{Paxton}, B., {Smolec}, R., {Schwab}, J., {et~al.} 2019, ApJS, 243, 10,
  \dodoi{10.3847/1538-4365/ab2241}

\bibitem[{{Portegies Zwart} \& {McMillan}(2002)}]{Portegies_Zwart2002}
{Portegies Zwart}, S.~F., \& {McMillan}, S. L.~W. 2002, ApJ, 576, 899,
  \dodoi{10.1086/341798}

\bibitem[{{Regan} {et~al.}(2014){Regan}, {Johansson}, \& {Wise}}]{Regan2014}
{Regan}, J.~A., {Johansson}, P.~H., \& {Wise}, J.~H. 2014, ApJ, 795, 137,
  \dodoi{10.1088/0004-637X/795/2/137}

\bibitem[{{Reinoso} {et~al.}(2018){Reinoso}, {Schleicher}, {Fellhauer},
  {Klessen}, \& {Boekholt}}]{Reinoso2018}
{Reinoso}, B., {Schleicher}, D.~R.~G., {Fellhauer}, M., {Klessen}, R.~S., \&
  {Boekholt}, T.~C.~N. 2018, A\&A, 614, A14,
  \dodoi{10.1051/0004-6361/201732224}

\bibitem[{{Sakurai} {et~al.}(2017){Sakurai}, {Yoshida}, {Fujii}, \&
  {Hirano}}]{Sakurai2017}
{Sakurai}, Y., {Yoshida}, N., {Fujii}, M.~S., \& {Hirano}, S. 2017, MNRAS, 472,
  1677, \dodoi{10.1093/mnras/stx2044}

\bibitem[{{Sat{\={o}}}(1966)}]{Sato1966}
{Sat{\={o}}}, H. 1966, Progress of Theoretical Physics, 35, 241,
  \dodoi{10.1143/PTP.35.241}

\bibitem[{{Schleicher} {et~al.}(2013){Schleicher}, {Palla}, {Ferrara}, {Galli},
  \& {Latif}}]{Schleicher2013}
{Schleicher}, D. R.~G., {Palla}, F., {Ferrara}, A., {Galli}, D., \& {Latif}, M.
  2013, A\&A, 558, A59, \dodoi{10.1051/0004-6361/201321949}

\bibitem[{{Schleicher} {et~al.}(2010){Schleicher}, {Spaans}, \&
  {Glover}}]{Schleicher2010a}
{Schleicher}, D. R.~G., {Spaans}, M., \& {Glover}, S. C.~O. 2010, ApJ, 712,
  L69, \dodoi{10.1088/2041-8205/712/1/L69}

\bibitem[{{Schwarzschild} \& {H{\"a}rm}(1959)}]{Schwarzschild1959}
{Schwarzschild}, M., \& {H{\"a}rm}, R. 1959, ApJ, 129, 637,
  \dodoi{10.1086/146662}

\bibitem[{{Shang} {et~al.}(2010){Shang}, {Bryan}, \& {Haiman}}]{Shang2010}
{Shang}, C., {Bryan}, G.~L., \& {Haiman}, Z. 2010, MNRAS, 402, 1249,
  \dodoi{10.1111/j.1365-2966.2009.15960.x}

\bibitem[{{Shapiro} \& {Teukolsky}(1983)}]{Shapiro1983}
{Shapiro}, S.~L., \& {Teukolsky}, S.~A. 1983, {Black holes, white dwarfs, and
  neutron stars : the physics of compact objects}

\bibitem[{{Shibata} \& {Shapiro}(2002)}]{Shibata2002}
{Shibata}, M., \& {Shapiro}, S.~L. 2002, ApJ, 572, L39, \dodoi{10.1086/341516}

\bibitem[{{Shiode} {et~al.}(2012){Shiode}, {Quataert}, \& {Arras}}]{Shiode2012}
{Shiode}, J.~H., {Quataert}, E., \& {Arras}, P. 2012, MNRAS, 423, 3397,
  \dodoi{10.1111/j.1365-2966.2012.21130.x}

\bibitem[{{Sonoi} \& {Umeda}(2012)}]{Sonoi2012}
{Sonoi}, T., \& {Umeda}, H. 2012, MNRAS, 421, L34,
  \dodoi{10.1111/j.1745-3933.2011.01201.x}

\bibitem[{{Stacy} {et~al.}(2016){Stacy}, {Bromm}, \& {Lee}}]{Stacy2016}
{Stacy}, A., {Bromm}, V., \& {Lee}, A.~T. 2016, MNRAS, 462, 1307,
  \dodoi{10.1093/mnras/stw1728}

\bibitem[{{Stacy} {et~al.}(2012){Stacy}, {Greif}, \& {Bromm}}]{Stacy2012}
{Stacy}, A., {Greif}, T.~H., \& {Bromm}, V. 2012, MNRAS, 422, 290,
  \dodoi{10.1111/j.1365-2966.2012.20605.x}

\bibitem[{{Sugimura} {et~al.}(2014){Sugimura}, {Omukai}, \&
  {Inoue}}]{Sugimura2014}
{Sugimura}, K., {Omukai}, K., \& {Inoue}, A.~K. 2014, MNRAS, 445, 544,
  \dodoi{10.1093/mnras/stu1778}

\bibitem[{{Susa} {et~al.}(2014){Susa}, {Hasegawa}, \& {Tominaga}}]{Susa2014}
{Susa}, H., {Hasegawa}, K., \& {Tominaga}, N. 2014, ApJ, 792, 32,
  \dodoi{10.1088/0004-637X/792/1/32}

\bibitem[{{Tagawa} {et~al.}(2020){Tagawa}, {Haiman}, \& {Kocsis}}]{Tagawa2020}
{Tagawa}, H., {Haiman}, Z., \& {Kocsis}, B. 2020, ApJ, 892, 36,
  \dodoi{10.3847/1538-4357/ab7922}

\bibitem[{{Uchida} {et~al.}(2017){Uchida}, {Shibata}, {Yoshida}, {Sekiguchi},
  \& {Umeda}}]{Uchida2017}
{Uchida}, H., {Shibata}, M., {Yoshida}, T., {Sekiguchi}, Y., \& {Umeda}, H.
  2017, PhRvD, 96, 083016, \dodoi{10.1103/PhysRevD.96.083016}

\bibitem[{{Umeda} {et~al.}(2016){Umeda}, {Hosokawa}, {Omukai}, \&
  {Yoshida}}]{Umeda2016}
{Umeda}, H., {Hosokawa}, T., {Omukai}, K., \& {Yoshida}, N. 2016, ApJ, 830,
  L34, \dodoi{10.3847/2041-8205/830/2/L34}

\bibitem[{{Unno} {et~al.}(1989){Unno}, {Osaki}, {Ando}, {Saio}, \&
  {Shibahashi}}]{Unno1989}
{Unno}, W., {Osaki}, Y., {Ando}, H., {Saio}, H., \& {Shibahashi}, H. 1989,
  {Nonradial oscillations of stars}

\bibitem[{{Venemans} {et~al.}(2013){Venemans}, {Findlay}, {Sutherland}, {De
  Rosa}, {McMahon}, {Simcoe}, {Gonz{\'a}lez-Solares}, {Kuijken}, \&
  {Lewis}}]{Venemans2013}
{Venemans}, B.~P., {Findlay}, J.~R., {Sutherland}, W.~J., {et~al.} 2013, ApJ,
  779, 24, \dodoi{10.1088/0004-637X/779/1/24}

\bibitem[{{Vink} {et~al.}(2001){Vink}, {de Koter}, \& {Lamers}}]{Vink2001}
{Vink}, J.~S., {de Koter}, A., \& {Lamers}, H.~J.~G.~L.~M. 2001, A\&A, 369,
  574, \dodoi{10.1051/0004-6361:20010127}

\bibitem[{{Volonteri}(2012)}]{Volonteri2012}
{Volonteri}, M. 2012, Science, 337, 544, \dodoi{10.1126/science.1220843}

\bibitem[{{Wolcott-Green} \& {Haiman}(2011)}]{Wolcott-Green2011}
{Wolcott-Green}, J., \& {Haiman}, Z. 2011, MNRAS, 412, 2603,
  \dodoi{10.1111/j.1365-2966.2010.18080.x}

\bibitem[{{Woosley}(2017)}]{Woosley2017}
{Woosley}, S.~E. 2017, ApJ, 836, 244, \dodoi{10.3847/1538-4357/836/2/244}

\bibitem[{{Woosley} {et~al.}(2002){Woosley}, {Heger}, \&
  {Weaver}}]{Woosley2002}
{Woosley}, S.~E., {Heger}, A., \& {Weaver}, T.~A. 2002, Reviews of Modern
  Physics, 74, 1015, \dodoi{10.1103/RevModPhys.74.1015}

\bibitem[{{Wu} {et~al.}(2015){Wu}, {Wang}, {Fan}, {Yi}, {Zuo}, {Bian}, {Jiang},
  {McGreer}, {Wang}, {Yang}, {Yang}, {Thompson}, \& {Beletsky}}]{Wu2015}
{Wu}, X.-B., {Wang}, F., {Fan}, X., {et~al.} 2015, Nature, 518, 512,
  \dodoi{10.1038/nature14241}

\bibitem[{{Yadav} {et~al.}(2018){Yadav}, {K{\"u}hnrich Biavatti}, \&
  {Glatzel}}]{Yadav2018}
{Yadav}, A.~P., {K{\"u}hnrich Biavatti}, S.~H., \& {Glatzel}, W. 2018, MNRAS,
  475, 4881, \dodoi{10.1093/mnras/sty092}

\bibitem[{{Yajima} \& {Khochfar}(2016)}]{Yajima2016}
{Yajima}, H., \& {Khochfar}, S. 2016, MNRAS, 457, 2423,
  \dodoi{10.1093/mnras/stw058}

\end{thebibliography}



\end{document}